\documentclass[apj]{emulateapj}
\usepackage[latin2]{inputenc}
\usepackage{amsmath}
\usepackage{url}
\usepackage{hyperref} 
\usepackage{natbib}
\usepackage{multirow}
\usepackage{epstopdf}
\usepackage{graphicx}
\usepackage{rotating}

\begin{document}

\title{Determining Type I\MakeLowercase{a} Supernovae Host galaxy extinction probabilities  and a statistical approach to estimating the absorption-to-reddening ratio $R_V$}
\author{
Aleksandar Cikota\altaffilmark{1,2,3}, 
Susana Deustua\altaffilmark{3}, 
Francine Marleau\altaffilmark{2} }
\altaffiltext{1}{European Southern Observatory, Karl-Schwarzschild-Str. 2, D-85748 Garching b. M\"{u}nchen, Germany; email: acikota@eso.org}
\altaffiltext{2}{Institute for Astro- and Particle Physics, University of Innsbruck, Technikerstrasse 25/8, A-6020 Innsbruck, Austria}
\altaffiltext{3}{Space Telescope Science Institute, 3700 San Martin Drive, Baltimore, MD 21218, USA}

\begin{abstract}
We investigate limits on the extinction values of Type Ia supernovae to statistically determine the most probable color excess, E(B-V), with  galactocentric distance, and use these statistics to determine the absorption-to-reddening ratio, $R_V$, for dust in the host galaxies. We determined pixel-based dust mass surface density maps for 59 galaxies from the Key Insight on Nearby Galaxies: a Far-Infrared Survey with \textit{Herschel} (KINGFISH, \citet{2011PASP..123.1347K}). We use Type Ia supernova spectral templates  \citep{Hsiao2007ApJ...663.1187H} to develop a Monte Carlo simulation of color excess E(B-V) with $R_V$ = 3.1 and investigate the color excess probabilities E(B-V) with projected radial galaxy center distance. Additionally, we tested our model using observed spectra of SN 1989B, SN 2002bo and SN 2006X, which occurred in three KINGFISH galaxies. Finally, we determined the most probable reddening for Sa-Sap, Sab-Sbp, Sbc-Scp, Scd-Sdm, S0 and Irregular galaxy classes as a function of $R/R_{25}$.  We find that the largest expected reddening probability are in Sab-Sb and Sbc-Sc galaxies, while S0 and Irregulars are very dust poor. We present a new approach for determining the absorption-to-reddening ratio $R_V$ using color excess probability functions, and find for a sample of 21 SNe Ia observed in Sab-Sbp galaxies, and 34 SNe in Sbc-Scp, an $R_V$ of 2.71 $\pm$ 1.58 and $R_V$ = 1.70 $\pm$ 0.38 respectively.
\end{abstract}


\maketitle

\section{Introduction}

 Because Type Ia supernovae (SNe Ia) are bright, they are good standard candles and probably the most accurate distance indicators on cosmological scales. Although SNe Ia are not equally bright there is a known correlation between their peak brightness and the width of their light curves \citep{Phillips1993ApJ...413L.105P}, which is used to standardize the "Branch" normal SNe Ia. However, there are SNe Ia that appear dimmer and redder than the Branch normals, either because they are intrinsically different, or because they suffer greater host galaxy extinction.  
 One of the largest sources of uncertainty in Type Ia SNe photometric measurements is the extinction due to the host galaxy, which affects the accuracy and precision of constructed Hubble diagrams. In turn, this limits the accuracy of the measurement of the dark energy parameters. \citet{Sullivan2006ApJ...648..868S} examined the effect of host galaxy morphology on the Hubble diagram of SNe Ia. They found that elliptical galaxy SNe Hubble diagram  had less scatter than spiral galaxy SNe.
 
   Understanding the effect of dust extinction on SNe Ia is essential for accurate measurement of cosmological parameters and the expansion history of the Universe \citep{1998AJ....116.1009R, 1999ApJ...517..565P}. 
Extinction $A_V$, is determined from the reddening law $E(B-V) = A_V/R_V$, where the color excess E(B-V) depends on the properties of dust. If these are uncertain and/or evolve with redshift, the  extinction, and thus the SNe Ia brightness, might be systematically wrong.
 
  Previous studies of the extinction from SNe Ia (summarized in Table~\ref{rvstudies}) yielded diverse values of the absorption to reddening ratio, $R_V$, ranging from $R_V$ = 1 to $R_V$ = 3.5. For comparison, the average value for the Milky Way is $R_V$ = 3.1. These studies used a variety of methods to calculate $R_V$, most of them are variants on multi-color light curve fitting.

\citet{1992ARA&A..30..359B} calculated $\langle R_B \rangle$ = 2.0 from measurements of six SNe Ia in the Virgo cluster, and $\langle R_B \rangle$ = 1.2 from three pairs of SNe which occurred in the same galaxies (note that $R_B$ = $R_V$ + 1). They also determined $R_B$ = 1.3 $\pm$ 0.2 from a least square solution for a sample of 17 nearby SNe Ia ($\lesssim$ 4000 $km$ $s^{-1}$) in the Hubble diagram, using the SNe Ia sample from \citet{1990AJ....100..530M}. \citet{1999AJ....118.1766P} developed a method to estimate the extinction of low redshift SNe Ia (0.01 $\lesssim$ z $\lesssim$ 0.1) based on the observational coincidence that the B-V evolution between 30 and 90 days after the peak luminosity in V is similar for all SNe Ia, regardless of light curve shape. They obtain $R_V$ = 3.5 $\pm$ 0.4 for a sample of 49 SNe Ia.

\citet{2004MNRAS.349.1344A} estimated $R_B$ = 3.5 from light curves of 73 SNe with z $\lesssim$ 0.1. \citet{2005ApJ...624..532R} calculated $R_V$ = 2.65 $\pm$ 0.15 for 111 SNe Ia with recession velocities between 3000 $km$ $s^{-1}$ and 20000 $km$ $s^{-1}$, exploiting the $M_B$(max) vs. E(B-V) correlation. 
\citet{1996ApJ...473..588R} found a value of $R_V$ = 2.55 $\pm$ 0.30, derived using a multi-color light-curve shape (MLCS) method for a sample of 20 Ia SNe with z $\lesssim$ 0.1.
\citet{2007ApJ...664L..13C} used four different light curve fitting packages, and determined $R_V$ $\sim$ 1 for a sample of 61 SNe with v $\lesssim$ 40000 $km$ $s^{-1}$. To explain their low $R_V$, \citet{2007ApJ...664L..13C} suggest that a more complicated model of intrinsic SN colors is required, which goes beyond single light-curve shape-color relation, or that dust in the host galaxies of the Ia SNe is quite different compared to Milky Way dust.

\citet{2009ApJ...700.1097H} combined the CfA3 SNe Ia \citep{2009ApJ...700..331H} with the UNION set \citep{2008ApJ...686..749K} to determine the equation of state parameter \textit{w}. They use four different light curve fitters and found lower Hubble residuals for $R_V$ = 1.7 compared to $R_V$ = 3.1, i.e. the higher $R_V$ value overestimates host galaxy extinction. 

In a different approach, \citet{2011ApJ...731..120M} constructed a statistical model for Type Ia SNe light curves from the visible through the infrared, and applied it to data of 127 SNe from three surveys (PAIRITEL, CfA3, Carnegie Supernova Project) and from the literature. They calculated $R_V$ $\approx$ 2.5 $-$ 2.9 for $A_V$ $\lesssim$ 0.4, while for higher extinctions, $A_V$ $\gtrsim$ 1, values of $R_V <$ 2 are calculated. \citet{2009ApJS..185...32K} determined $R_V$ = 2.18 $\pm$ 0.14(stat) $\pm$ 0.48(syst) by matching observed and predicted SN Ia colors for 103 SNe with 0.04 $<$ z $<$ 0.42. \citet{2010AJ....139..120F} found $R_V$ $\approx$ 1.7 when using 17 low-redshift (z $<$ 0.08) SNe Ia monitored by the Carnegie Supernova Project, but obtain $R_V$ $\approx$ 3.2 when two highly reddened SNe are excluded.

\citet{2010ApJ...722..566L} from a sample of 361 SDSS-II SNe, z $<$ 0.21, utilizing two light curve methods, found that SNe Ia in passive host galaxies favor a dust law of $R_V$ = 1.0 $\pm$ 0.2, while SNe Ia in star-forming hosts require $R_V$ = $1.8^{+0.2}_{-0.4}$. 

\citet{Nobili2008A&A...487...19N} found that SNe Ia color depends on the light curve shape, such that SNe Ia with fainter, narrower light curves are redder than those with brighter, broader light curves (cf. {\citet{Phillips1993ApJ...413L.105P}}, {\citet{1996ApJ...473...88R}}, {\citet{1999AJ....118.1766P}}, {\citet{2002PASP..114..803N}} and {\citet{2003A&A...404..901N}}). They correct SNe light curves for this intrinsic color difference, and then derive host galaxy reddening. They obtain $R_V$ = 1.75 $\pm$ 0.27 for 80 low redshift ($\lesssim$ 0.1) SNe Ia with $E(B-V) \le 0.7$ mag, but find that a subset of 69 SNe that have modest reddening, $E(B-V) < 0.25$ mag, have significantly smaller $R_V$ $\sim$ 1. 

Using a Monte Carlo simulation of circumstellar dust around the supernova location, \citet{Goobar2008ApJ...686L.103G} determined $R_V$ $\sim$ 1.5 $-$ 2.5 for SNe Ia. One of Goobar's motivations in undertaking this study was the observed steeper dependence with wavelength of the total to selective extinction seen in SNe Ia by e.g. \citet{Wang2005ApJ...635L..33W}. Using Draine's Milky Way dust models \citep{2003ApJ...598.1026D} and the \citet*{Weingartner2001ApJ...548..296W} Large Magellanic Clouds dust properties in his simulation, Goobar finds that a simple power law fits the resulting extinction.\\

   Given the apparent inconsistent results in estimating host galaxy extinction from SN color observations, we decided to investigate this problem using a different approach. 
Our investigation instead concentrates on host galaxy properties, using data obtained in a systematic and consistent way for nearby galaxies to measure the mass and distribution of interstellar medium components (dust), and thence estimate the extinction. The principal goal is to place limits on the uncertainties of SNe Ia extinction values, to estimate the range of host galaxy extinction, i.e. the most probable color excess, in a galaxy sample, and then apply these statistics to a sample of observed Ia SNe in order to infer $R_V$. 

	We opted to use the KINGFISH galaxy sample (Key Insight on Nearby Galaxies: a Far-Infrared Survey with \textit{Herschel}, \citealp{2011PASP..123.1347K}), which when combined with SINGS (Spitzer Infrared Nearby Galaxies Survey (SINGS, \citealp{Kennicutt2003PASP..115..928K}), is one of the most complete multi-band surveys of nearby galaxies.  Further, \citet{Draine2007ApJ...663..866D},  \citet{Gordon2008} and \citet{2011ApJ...738...89S} have estimated the dust mass of these galaxies. Therefore, these galaxies are an ideal laboratory to explore the effects of dust extinction on SNe Ia. 
    
    We first determine the dust density on a per pixel base, then we look at the change in SNe Ia colors with galactocentric distance due to dust extinction.
    
For galaxies with known SNe Ia, we compare the extincted spectrum template at the position of the historical supernova to the observed spectra.
Finally we present color excess E(B-V) probabilities as functions of galactocentric distance for SNe Ia in different morphological host galaxy types, and use those reddening probability functions to estimate the absorption-to-reddening ratio, $R_V$, of dust in SNe Ia host galaxies.

	In $\S$2 we describe the data, the galaxy dust mass surface density maps, the extinction model, and the spectral templates of SNe Ia. In $\S$3 we describe our Monte Carlo simulation on host galaxy extinction and compare the model to individual Ia observations. In $\S$4 we show and discuss the simulation results, apply them for $R_V$ determination of a Ia SNe sample and discuss the uncertainties. In $\S$5 we summarize the results and conclusions.

\begin{table*}
\caption{$R_V$ results of earlier studies}
\vspace{-5mm}
\label{rvstudies}
\begin{center}
\begin{tabular}{l l l l }
\hline\hline
\sc Reference  & \sc No. of SNe & \sc Velocity or redshift & \sc Absorption-to-reddening ratio \\
\hline\hline
{\citet{1992ARA&A..30..359B}} & 6 & Nearby (Virgo cluster) 			& $\langle R_B \rangle$ = 2.0 \\ 
                              & 3 pairs &  Nearby 				& $\langle R_B \rangle$ = 1.2 \\ 
                              & 17 & \textit{v} $\lesssim$ 4000 km $s^{-1}$ & $R_B$ = 1.3 $\pm$ 0.2 \\ 
{\citet{1996ApJ...473..588R}} & 20 &  z $\lesssim$ 0.1& $R_V$ = 2.55 $\pm$ 0.30 \\
{\citet{1999AJ....118.1766P}} & 49  & 0.01 $\lesssim$ z $\lesssim$ 0.1 & $R_V$ = 3.5 $\pm$ 0.4\\
{\citet{2004MNRAS.349.1344A}} & 73 &  z $\lesssim$0.1 & $R_B$ = 3.5 \\
{\citet{2005ApJ...624..532R}} & 111  & 3000 km $s^{-1}$ $<\textit{v}_{CMB}<$ 20000 km $s^{-1}$ &$R_V$ = 2.65 $\pm$ 0.15  \\
{\citet{2007ApJ...664L..13C}} & 28-61 & z $\lesssim$ 0.13  & $R_V$ $\sim$ 1\\
{\citet{2009ApJ...700.1097H}} & 70-203 & 0.01 $\lesssim$ z $\lesssim$ 1.1 & $R_V \approx$ 1.7 \\
{\citet{2011ApJ...731..120M}} &127 & z $\lesssim$ 0.05 & $R_V$ $\approx$ 2.5-2.9 for $A_V$ $\lesssim$ 0.4\\ 
				              & & & $R_V <$ 2 for $A_V$ $\gtrsim$ 1\\
{\citet{Nobili2008A&A...487...19N}}   &  80 & z $\lesssim$  0.1&  $R_V = 1.75 \pm 0.27$ for $E(B-V)  \le 0.7$ mag\\
                                    & &   & $R_V \sim 1$ for $E(B-V) < 0.25$ mag\\
{\citet{2009ApJS..185...32K}} &103 & 0.04 $<$ z $<$ 0.42 & $R_V$ = 2.18 $\pm$ 0.14(stat) $\pm$ 0.48(syst) \\
{\citet{2010AJ....139..120F}} & 17 & z $<$ 0.08 & $R_V$ $\approx$ 1.7 \\
							  & &	&	$R_V \approx$ 3.2 when two reddened SNe excluded\\
{\citet{2010ApJ...722..566L}} &361 & z $<$ 0.21 & $R_V$ = 1.0 $\pm$ 0.2 in passive host \\
				 			&	&	&	$R_V$ = $1.8^{+0.2}_{-0.4}$ in star-forming hosts\\
{\citet{Goobar2008ApJ...686L.103G}}&  \multicolumn{2}{l}{\sc Monte Carlo simulation} & $R_V$ $\sim$ 1.5 - 2.5 \\                          
                              
\hline\hline
\end{tabular}
\end{center}
\end{table*}

\section{Data and models}
\label{data and models}

\subsection{KINGFISH sample}
\label{kfsample}
The KINGFISH project is an imaging and spectroscopic survey of 61 nearby (d $<$ 30 Mpc) galaxies, including 59 galaxies from the SINGS project, for which dust mass estimates have been determined (\citet{Draine2007ApJ...663..866D, 2011ApJ...738...89S}). The galaxy sample covers the full range of integrated properties and local interstellar medium environments found in the nearby Universe.
Compared to Spitzer, whose limited wavelength coverage makes it difficult to separate dust temperature distributions from grain emissivity functions, Herschel's deep, submillimeter imaging capability at 250, 350, and 500 $\mu$m with the SPIRE (Spectral and Photometric Imaging Receiver) instrument, enables the direct detection of cool dust and constrains the Rayleigh-Jeans region of the main dust emission components \citep{2011PASP..123.1347K}.

For our purposes, our sample consists of 59 KINGFISH galaxies (Table~\ref{Kingfishgalaxies}). We excluded NGC 1404 and DDO 154 because they lack SPIRE fluxes. The galaxies are grouped according to their morphology: Sa-Sap, Sab-Sbp, Sbc-Scp, Scd-Sdm, S0 and Irregulars, containing 10, 8, 11, 11, 8 and 9 galaxies in the group respectively. NGC3265 is the only elliptical and NGC4625 is a dwarf spiral (SABm).  
For each galaxy, we created dust mass surface density maps on a pixel by pixel basis. The redshift-independent galaxy distances and heliocentric radial velocities in Table~\ref{Kingfishgalaxies} are from \citet{2011PASP..123.1347K}. The de Vaucouleur radius $R_{25}$ are from the RC3 \citep{1991rc3..book.....D}. Galaxy types and dust temperatures are from \citet{2011ApJ...738...89S}.

\begin{table*}
\caption{Sample of the KINGFISH galaxies}
\label{Kingfishgalaxies}
\vspace{-5mm}
\begin{center}
\begin{tabular}{l|llcccll|ccc}
\hline\hline
\multirow{5}{*}{\begin{sideways}Group\end{sideways}} & & &&&&& \textit{This work} & $Skibba$ & $Draine$ & $Gordon$ \\
& \multicolumn{1}{c}{\sc Galaxy} & \multicolumn{1}{c}{\sc Type} & \multicolumn{1}{c}{$v$}      & \multicolumn{1}{c}{Distance} & \multicolumn{1}{c}{\sc $R_{25}$}        & \multicolumn{1}{c}{\sc T$_{dust}$} & \multicolumn{1}{c}{\sc $logM_{dust}$} & \multicolumn{1}{|c}{\sc $logM_{dust}$} & \multicolumn{1}{c}{\sc $logM_{dust}$} & \multicolumn{1}{c}{\sc $logM_{dust}$} \\ 
&           &          &  \multicolumn{1}{c}{\sc ($km s^{-1}$)} & \multicolumn{1}{c}{\sc ($Mpc$)}    & \multicolumn{1}{c}{\sc ($arcmin$)}   &  \multicolumn{1}{c}{\sc (K) }      & \multicolumn{1}{c}{\sc ($log M_{\odot}$)} & \multicolumn{1}{|c}{($log M_{\odot}$)}& ($log M_{\odot}$)& ($log M_{\odot}$) \\            
& \multicolumn{1}{c}{\sc (1)} & \multicolumn{1}{c}{\sc(2)} &  \multicolumn{1}{c}{\sc (3)} & \multicolumn{1}{c}{\sc (4)} & \multicolumn{1}{c}{\sc(5)} & \multicolumn{1}{c}{\sc(6)} & \multicolumn{1}{c}{\sc(7)} & \multicolumn{1}{|c}{\sc(8)} & \multicolumn{1}{c}{\sc(9)}& \multicolumn{1}{c}{\sc(10)} \\ 
\hline\hline
\multirow{10}{*}{\begin{sideways}Sa-Sap\end{sideways}} & NGC 1482 &       Sa & 1655   &    22.6 &        1.3 &    31.8 $\pm$ 0.9  &    7.13 $\pm$ 0.04   &    7.13 $\pm$    0.08 &  7.47 &  7.43 \\
& NGC 1512 &      SBa &  896   &   14.35 &        4.5 &    20.9 $\pm$ 0.8  &    7.11 $\pm$ 0.17   &     7.0 $\pm$    0.08 &  7.21 &  6.96 \\
& NGC 2798 &    SABap & 1726   &    25.8 &        1.3 &    34.9 $\pm$ 1.1  &    6.87 $\pm$ 0.07   &    6.83 $\pm$    0.08 &  7.29 &  7.22 \\
& NGC 2841 &     SABa &  638   &    14.1 &        4.1 &    22.1 $\pm$ 0.4  &    7.44 $\pm$ 0.07   &    7.34 $\pm$    0.08 &  7.74 &  7.44 \\
& NGC 3351 &      SBa &  778   &     9.8 &        3.7 &    25.6 $\pm$ 0.6  &    6.91 $\pm$ 0.09   &    6.87 $\pm$    0.08 &  7.46 &  7.32 \\
& NGC 3190 &     SAap & 1271   &    19.3 &        2.2 &    25.2 $\pm$ 0.5  &    6.97 $\pm$ 0.12   &    6.89 $\pm$    0.08 &  7.19 &  7.11 \\
& NGC 4579 &      SBa & 1519   &    15.3 &        2.9 &    23.4 $\pm$ 0.5  &    7.19 $\pm$ 0.06   &    7.12 $\pm$    0.08 &  8.18 &  7.77 \\
& NGC 4594 &      SAa & 1091   &     9.4 &        4.4 &    22.1 $\pm$ 0.4  &    6.99 $\pm$ 0.07   &    6.91 $\pm$    0.08 &  7.56 &  7.56 \\
& NGC 4725 &     SABa & 1206   &    12.7 &        5.4 &    21.1 $\pm$ 0.4  &    7.41 $\pm$ 0.09   &    7.34 $\pm$    0.08 &   8.2 &  7.88 \\
& NGC 4736 &     SABa &  308   &    4.66 &        5.6 &    29.3 $\pm$ 0.8  &    6.55 $\pm$ 0.09   &    6.52 $\pm$    0.08 &  7.11 &  6.97 \\
\cline{1-1}
\multirow{8}{*}{\begin{sideways}Sab-Sbp\end{sideways}} & NGC 1097 &    SBabp & 1275   &   19.09 &        4.7 &    26.2 $\pm$ 0.6  &    7.86 $\pm$ 0.06   &     7.8 $\pm$    0.08 &  8.37 &  8.05 \\
& NGC 2146 &    SBabp &   893  &    17.2 &        3.0 &    37.4 $\pm$ 1.2  &    7.45 $\pm$ 0.04   &    7.36 $\pm$    0.08 &  \ldots &  \ldots \\
& NGC 3049 &     SBab & 1494   &    19.2 &        1.1 &    27.5 $\pm$ 0.7  &    6.55 $\pm$ 0.06   &    6.45 $\pm$    0.08 &  6.74 &  6.62 \\
& NGC 3627 &     SBbp &  727   &    10.3 &        4.6 &    27.2 $\pm$ 0.7  &    7.38 $\pm$ 0.05   &    7.32 $\pm$    0.08 &  7.69 &  7.61 \\
& NGC 4569 &    SABab & -235   &    15.3 &        4.8 &    24.0 $\pm$ 0.5  &    7.22 $\pm$ 0.17   &    7.16 $\pm$    0.08 &  7.75 &  7.67 \\
& NGC 4826 &     SAab &  408   &    5.57 &        5.0 &    29.1 $\pm$ 0.8  &     6.5 $\pm$ 0.09   &    6.38 $\pm$    0.08 &  6.89 &  6.77 \\
& NGC 5713 &    SBabp & 1883   &   21.37 &        1.4 &    30.0 $\pm$ 0.8  &    7.15 $\pm$ 0.04   &    7.07 $\pm$    0.08 &  7.95 &   7.7 \\
& NGC 7331 &      SAb &  816   &    14.9 &        5.2 &    26.1 $\pm$ 0.6  &    7.82 $\pm$ 0.05   &    7.71 $\pm$    0.08 &  8.05 &  7.99 \\
\cline{1-1}
\multirow{11}{*}{\begin{sideways}Sbc-Scp\end{sideways}} & NGC 628 &      SAc &  657   &     7.3 &        5.2 &    24.0 $\pm$ 0.6  &    7.09 $\pm$ 0.07   &    7.03 $\pm$    0.08 &  8.02 &  7.58 \\
& NGC 3184 &     SAbc &  592   &     8.7 &        3.7 &    23.4 $\pm$ 0.5  &    6.95 $\pm$ 0.06   &     6.9 $\pm$    0.08 &   7.7 &  7.15 \\
& NGC 3198 &    SABbc &  663   &    14.5 &        4.3 &    23.6 $\pm$ 0.5  &    7.26 $\pm$ 0.09   &    7.18 $\pm$    0.08 &  7.42 &  7.07 \\
& NGC 3521 &    SABbc &  805   &   12.44 &        5.5 &    24.9 $\pm$ 0.6  &    7.71 $\pm$ 0.05   &    7.63 $\pm$    0.08 &  7.83 &   7.7 \\
& NGC 3938 &      SAc &  809   &    12.1 &        2.7 &    24.8 $\pm$ 0.5  &    7.03 $\pm$ 0.05   &    6.94 $\pm$    0.08 &  7.69 &  7.37 \\
& NGC 4254 &     SAcp & 2407   &    15.3 &        2.7 &    25.5 $\pm$ 0.5  &    7.57 $\pm$ 0.02   &    7.56 $\pm$    0.08 &  8.55 &  8.16 \\
& NGC 4321 &    SABbc & 1571   &    15.3 &        3.7 &    24.4 $\pm$ 0.5  &    7.64 $\pm$ 0.03   &    7.61 $\pm$    0.08 &  8.57 &  8.22 \\
& NGC 4536 &    SABbc & 1808   &    15.3 &        3.8 &    26.9 $\pm$ 0.6  &    7.31 $\pm$ 0.07   &    7.28 $\pm$    0.08 &   7.8 &  7.85 \\
& NGC 5055 &     SAbc &  504   &   10.16 &        6.3 &    24.1 $\pm$ 0.5  &    7.69 $\pm$ 0.05   &    7.61 $\pm$    0.08 &  8.19 &  7.87 \\
& NGC 5457 &       Sc &   241  &     7.1 &       14.4 &    24.3 $\pm$ 0.6  &    7.64 $\pm$ 0.1    &    7.52 $\pm$    0.08 &  \ldots &  \ldots \\
& NGC 7793 &      SAc &  230   &    3.91 &        4.7 &    24.1 $\pm$ 0.6  &    6.58 $\pm$ 0.05   &    6.51 $\pm$    0.08 &  6.92 &  6.52 \\
\cline{1-1}
\multirow{11}{*}{\begin{sideways}Scd-Sdm\end{sideways}} &  IC 342 &    SABcd &    31  &    3.28 &       10.7 &    24.1 $\pm$ 0.6  &    7.25 $\pm$ 0.03   &    7.27 $\pm$    0.05 &  \ldots &  \ldots \\
& NGC 337 &   SABcdp & 1650   &    22.9 &        1.4 &    28.1 $\pm$ 0.7  &    7.13 $\pm$ 0.05   &    7.07 $\pm$    0.08 &  7.65 &  7.38 \\
& NGC 925 &     SABd &  553   &    9.04 &        5.2 &    23.7 $\pm$ 0.5  &    7.06 $\pm$ 0.08   &    6.98 $\pm$    0.08 &  7.35 &  7.06 \\
& NGC 2976 &     SABd &    3   &     3.6 &        2.9 &    25.9 $\pm$ 0.7  &    6.05 $\pm$ 0.06   &    5.97 $\pm$    0.08 &  6.34 &   6.2 \\
& NGC 3621 &      SAd &  727   &     6.9 &        6.2 &    25.4 $\pm$ 0.6  &    7.09 $\pm$ 0.09   &    6.97 $\pm$    0.08 &  7.38 &   7.2 \\
& NGC 4236 &     SBdm &    0   &     3.6 &       10.9 &    25.0 $\pm$ 0.7  &    6.23 $\pm$ 0.31   &    5.83 $\pm$    0.08 &  6.15 &  5.79 \\
& NGC 4559 &     SBcd &  816   &    8.45 &        5.4 &    24.5 $\pm$ 0.5  &    6.94 $\pm$ 0.06   &    6.83 $\pm$    0.08 &  7.57 &  7.24 \\
& NGC 4631 &      SBd &  606   &    7.62 &        7.7 &    27.7 $\pm$ 0.8  &    7.35 $\pm$ 0.09   &    7.26 $\pm$    0.08 &  8.11 &  7.68 \\
& NGC 5398 &     SBdm & 1216   &    8.33 &        1.4 &    27.3 $\pm$ 0.7  &    5.71 $\pm$ 0.15   &    5.59 $\pm$    0.08 &  \ldots &  \ldots \\
& NGC 5474 &     SAcd &  273   &     6.8 &        2.4 &    24.6 $\pm$ 0.6  &    6.06 $\pm$ 0.14   &     6.0 $\pm$    0.08 &  6.39 &  6.06 \\
& NGC 6946 &    SABcd &   48   &     6.8 &        5.7 &    26.0 $\pm$ 0.6  &    7.49 $\pm$ 0.03   &    7.47 $\pm$    0.08 &  7.74 &  7.52 \\
\cline{1-1}
\multirow{8}{*}{\begin{sideways}S0\end{sideways}} & NGC 584 &     SAB0 & 1854   &    20.8 &        2.1 &    24.5 $\pm$ 0.6  &    6.27 $\pm$ 0.77   &    5.58 $\pm$    0.15 &  \ldots &  \ldots \\
& NGC 855 &      SA0 &  610   &    9.73 &        1.3 &    28.5 $\pm$ 0.9  &    5.56 $\pm$ 0.27   &    5.49 $\pm$    0.08 &  5.69 &  5.57 \\
& NGC 1266 &      SB0 & 2194   &    30.6 &        0.8 &    36.0 $\pm$ 1.0  &     6.7 $\pm$ 0.06   &    6.66 $\pm$    0.08 &  7.05 &  7.27 \\
& NGC 1291 &     SAB0 &  839   &    10.4 &        4.9 &    22.4 $\pm$ 0.5  &     6.8 $\pm$ 0.32   &    6.76 $\pm$    0.08 &  7.34 &  7.29 \\
& NGC 1316 &     SAB0 & 1760   &    20.1 &        6.0 &    26.8 $\pm$ 0.7  &    7.04 $\pm$ 1.64   &    6.79 $\pm$    0.08 &  7.63 &  7.67 \\
& NGC 1377 &       S0 & 1792   &    24.6 &        0.9 &    43.5 $\pm$ 1.8  &    5.95 $\pm$ 0.16   &    5.78 $\pm$    0.09 &  \ldots &  \ldots \\
& NGC 3773 &      SA0 &  987   &    12.4 &        0.6 &    30.2 $\pm$ 0.8  &    5.46 $\pm$ 0.09   &    5.44 $\pm$    0.08 &   5.9 &   6.0 \\
& NGC 5866 &       S0 &  692   &    15.3 &        2.4 &    27.9 $\pm$ 0.7  &    6.68 $\pm$ 0.17   &    6.57 $\pm$    0.08 &  6.65 &  6.84 \\
\cline{1-1}
\multirow{9}{*}{\begin{sideways}Irregulars\end{sideways}} &  DDO 53 &       Im &   19   &     3.6 &        0.8 &    30.5 $\pm$ 0.9  &    3.87 $\pm$ 0.33   &    4.01 $\pm$    0.10 &   4.0 &  4.35 \\
&  DDO 165 &       Im &   37   &     3.6 &        1.7 &    23.5 $\pm$ 1.1  &    4.48 $\pm$ 0.33   &    4.19 $\pm$    0.10 &  \ldots &  \ldots \\
&     Ho I &     IABm &  143   &     3.6 &        1.8 &    26.2 $\pm$ 0.9  &    4.42 $\pm$ 0.42   &    4.54 $\pm$    0.08 &  4.83 &  4.61 \\
&    Ho II &       Im &  157   &     3.6 &        4.0 &    36.5 $\pm$ 1.1  &    5.08 $\pm$ 0.89   &    4.05 $\pm$    0.20 &  5.07 &  5.38 \\
&  IC 2574 &      IBm &   57   &     3.6 &        6.6 &    25.9 $\pm$ 0.6  &    5.93 $\pm$ 0.32   &    5.57 $\pm$    0.08 &  5.86 &  6.04 \\
&  M81dwB &       Im &  350   &     3.6 &        0.4 &    25.0 $\pm$ 0.7  &    3.61 $\pm$ 0.28   &    4.06 $\pm$    0.09 &  \ldots &  \ldots \\
& NGC 2915 &       I0 &  468   &    3.78 &        0.9 &    28.9 $\pm$ 0.9  &    4.61 $\pm$ 0.16   &    4.59 $\pm$    0.08 &  4.14 &  3.77 \\
& NGC 3077 &      I0p &    14  &     3.6 &        2.7 &    30.1 $\pm$ 0.9  &    5.67 $\pm$ 0.07   &    5.52 $\pm$    0.08 &  \ldots &  \ldots \\
& NGC 5408 &      IBm &  509   &     4.8 &        0.8 &    25.7 $\pm$ 1.1  &    4.66 $\pm$ 0.23   &    4.68 $\pm$    0.08 &  4.67 &  4.09 \\
\cline{1-1}
& NGC 3265 &        E & 1421   &    19.6 &        0.6 &    31.8 $\pm$ 0.9  &    5.92 $\pm$ 0.07   &     6.0 $\pm$    0.08 &  6.17 &  6.28 \\
& NGC 4625 &     SABm &  609   &     9.3 &        1.1 &    24.8 $\pm$ 0.6  &    5.84 $\pm$ 0.08   &    5.89 $\pm$    0.08 &  6.35 &  6.18 \\
\hline\hline
\end{tabular}
\end{center}
\vspace{-2mm}
\textbf{Notes.} The galaxy morphological types (2), heliocentric radial velocities (3) and redshift-independent galaxy distances (4) were taken from Table 1 in \citet{2011ApJ...738...89S}. They obtained the morphological types from \citet{2010ApJS..190..147B} and \cite{Kennicutt2003PASP..115..928K}.
The de Vaucouleur's radii $R_{25}$ (5) were calculated from RC3 $D_{25}$ diameters \citep{1991rc3..book.....D}. The global dust temperatures (6) were determined by \citet{2011ApJ...738...89S}. The dust mass in column (7) is our integrated dust mass within 1 $R_{25}$. Dust masses in column (8) were determined by \citet{2011ApJ...738...89S}, in column (9) by \citet{Draine2007ApJ...663..866D} for SINGS, and in column (10) by \citet{Gordon2008} for SINGS, and are listed for comparison purposes. 
\end{table*}

\subsection{Dust mass maps}

\citet{2011ApJ...738...89S} calculate the total dust mass for each KINGFISH galaxy under the assumption that the dust radiates as a blackbody:
\begin{equation}
\label{eqdmass}
	\mathrm{M_{dust} = \frac{\textit{f}_\lambda 4 \pi D^2}{\kappa_{abs,\lambda}4\pi B_\lambda(T_{dust})} },
\end{equation}
where $\textit{f}_\lambda$ = $\textit{f}_\nu$c/$\lambda^2$ is the flux density, D, the distance to the galaxy, $\kappa_{abs}$ is the mass absorption coefficient and $B_\lambda$ = 2c$k_B$T/$\lambda^4$ is the Planck function in the Rayleigh-Jeans limit. They determine the dust temperature by fitting a single temperature blackbody curve to the \textit{Spitzer} MIPS and \textit{Herschel} SPIRE FIR and submm flux densities, assuming a dust emissivity $\epsilon \propto \lambda^{-\beta}B_\lambda(T_{dust})$, with $\beta$=1.5. Temperature uncertainties are determined from a Monte Carlo analysis that includes the flux errors. 
Dust masses are calculated using the standard Milky Way dust model with $R_V$ = 3.1 \citep{Weingartner2001ApJ...548..296W}, and $\kappa_{abs, 500\mu m}$ = 0.95 $cm^2 g^{-1}$ \citep{2003ARA&A..41..241D}. Dust masses are computed at 500 $\mu$m to minimize the temperature dependence, although the flux uncertainties are larger than at short wavelengths, where the estimated masses are lower \citep{2011ApJ...738...89S}.\\

  We create dust mass surface density maps on a pixel-by-pixel basis using equation~\ref{eqdmass}, the \citet{2011ApJ...738...89S} global dust temperatures and the Herschel Interactive Processing Environment (HIPE) processed, background subtracted 500 $\mu$m maps \citep{2011PASP..123.1347K}. SPIRE maps have wavelength dependent  pixel scales; at 500 $\mu$m it is 14 arcsec/pixel. We convert map fluxes from MJy/sr to MJy/parsec using the distances in \citet{2011ApJ...738...89S}.

As a cross-check, we integrated the mass over all the pixels within the de Vaucouleur's radius $R_{25}$, and compared them to the \citet{2011ApJ...738...89S} values. Since we use the same method, we would expect our estimates to be similar. We find that the median value of the ratios of our integrated mass determination to Skibba's is 1.18, dominated by  NGC 4236 and NGC 0584 whose estimated masses are 4 and 10 times larger, respectively. This result is mainly due to the choice of aperture. \citet{2011ApJ...738...89S} used the 3.6 $\mu$m images to create elliptical apertures encompassing the optical and infrared emission of the galaxy, while we integrated the mass within a circular aperture of $R_{25}$. For face-on galaxies our integrated mass is within a few percent of that calculated by \citet{2011ApJ...738...89S}; dust masses for NGC1482, NGC2915, NGC3773, NGC4254, NGC5408, NGC6946, IC0342 are within 5$\%$ of Skibba's masses. 
Another reason for the slightly different masses may be that they used differently processed Herschel maps, with slightly different background subtraction or calibration. The HIPE reduced maps have 15-20$\%$ negative pixel values, which we set to zero.

   Our results are in column 7 of Table~\ref{Kingfishgalaxies}. Note that errors are statistical due to the dust temperature and 500 $\mu$m flux uncertainties, and do not include systematic uncertainties.
In column 9 of Table~\ref{Kingfishgalaxies} we list dust masses determined by \citet{Draine2007ApJ...663..866D} from the SINGS data, which are $\sim$2-3 times larger than larger than Skibba's. These were calculated assuming a multi-temperature dust model consisting of a mixture of different grain types with a distribution of grain sizes.

\begin{figure}
\begin{center}
\includegraphics[trim=7mm 0mm 7mm 0mm, width=9cm, clip=true]{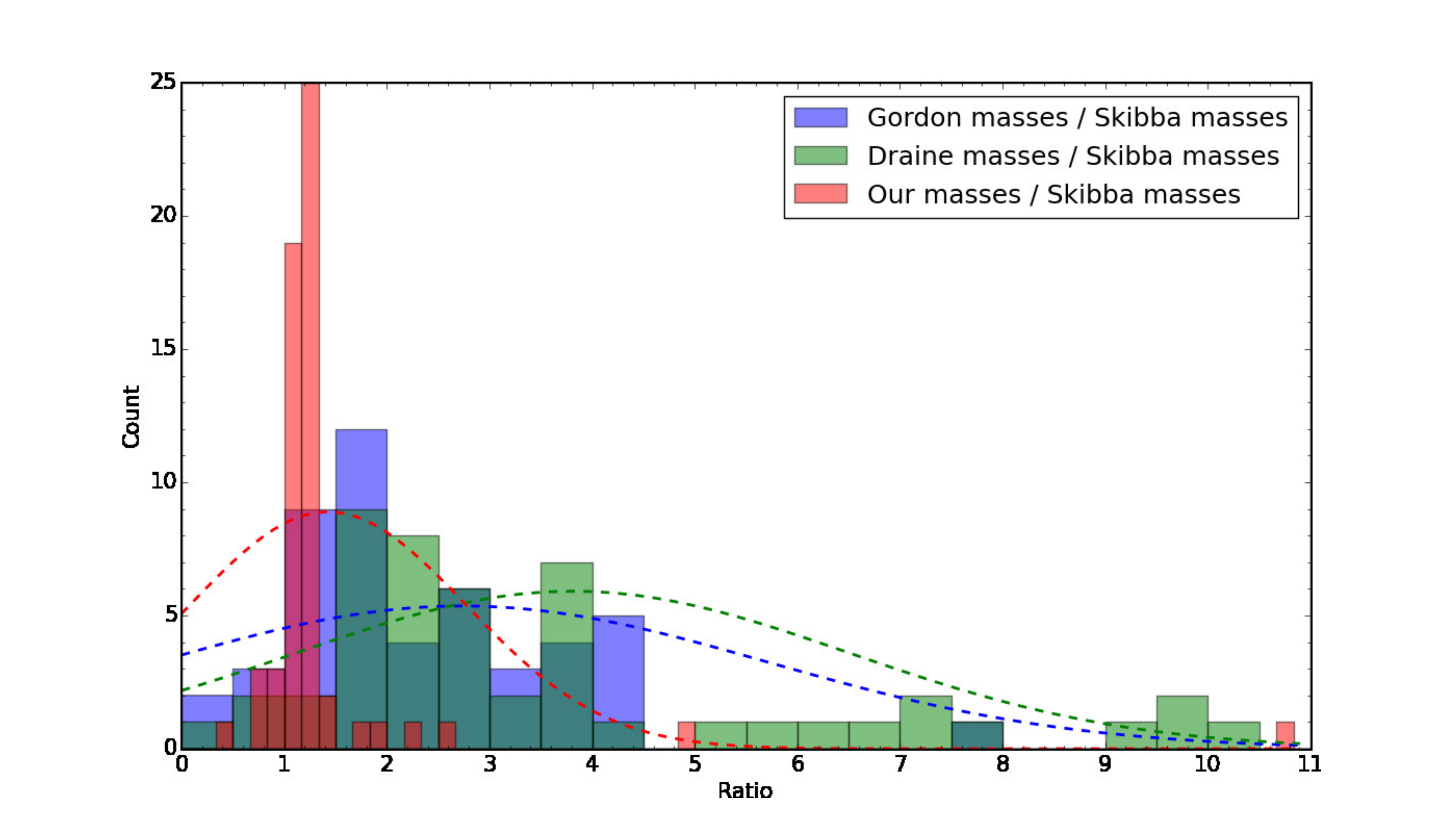}
\vspace{-5mm}
\caption{Ratios of Gordon, Draine and our dust masses to Skibba's (blue, green and red bars respectively). The dashed blue, green and red lines are gaussian fits of the corresponding histograms. Average dust mass ratios are  $Gordon/Skibba$ = 2.73 with $\sigma$ = 2.98, for  $Draine/Skibba$ = 3.81 with $\sigma$ = 2.70, and for $This Work/Skibba$ = 1.42 with $\sigma$ = 1.34.}
\label{fig:masshisto}
\end{center}
\end{figure}

In column 10 we list dust masses for SINGS galaxies estimated by \citet{Gordon2008} from the 70  $\mu$m to 160 $\mu$m flux ratio, assuming the dust radiates as a black body whose emissivity is proportional to $\lambda^{-2}$. He used a single dust temperature, estimated from the flux ratio, which combined with the measured 160 $\mu$m surface brightness and a simple, homogeneous dust grain model consisting of 0.1 $\mu$m silicon grains with the standard grain emissivity at 160 $\mu$m, provides the dust mass (cf. \citet{2010A&A...518L..89G}). Dust masses obtained in this manner are $\sim$ 2 times larger than ours, but smaller than Draine et al.'s (2007). 
Ratios of Gordon, Draine and our dust masses to Skibba dust masses are shown in Figure~\ref{fig:masshisto}. The influence of the total dust mass uncertainty on the results is discussed in $\S$4.3.

\subsection{Dust model and extinction law}

The amount of extinction depends on wavelength, the dust mass surface density, and the dust model. Extinction can be calculated from 
\begin{equation}
\label{eqA}
	\mathrm{A_\lambda = [(A_\lambda/N_H)(M_H/M_{dust})/m_H]\sigma_{dust}},
\end{equation}
and the extinction in V is
\begin{center}
$A_V=[(A_V/N_H)(M_H/M_{dust})/m_H]\sigma_{dust}$,
\end{center}
where $\sigma_{dust}$ is the dust mass surface density, $N_H$ the H column density, $m_H$ is the mass of a hydrogen atom, and $A_V$/$N_H$ is the attenuation per unit column density.
The Milky Way $R_V$ = 3.1 dust models of \citet{Weingartner2001ApJ...548..296W} have
 $A_V$/$N_H$ = 5.3$\times$10$^{-22}$ mag cm$^2$ H$^{-1}$, and a gas-to-dust mass ratio $M_H$/$M_{dust}$ = 98.0392 (see Table 3 in \citet{Draine2007ApJ...663..866D}).

We apply the \citet*{Cardelli1989ApJ...345..245C} (hereafter CCM) extinction law to calculate the extinction curve:
\begin{equation}
	\mathrm{\dfrac{A(\lambda)}{A(V)} = a(x)+b(x)/R_V}
\end{equation}
where a(x) and b(x) are wavelength dependent coefficients given in CCM. \\
At infrared wavelengths, 0.3 $\micron^{-1} \leqslant$ x $\leqslant 1.1 \micron^{-1}$:
\begin{equation*}
\mathrm{a(x)=0.574x^{1.61}} \qquad 
and \qquad 
\mathrm{b(x)=-0.527x^{1.61}}
\end{equation*}
and in the visible/NIR, 1.1 $\micron^{-1} \leqslant$ x $\leqslant 3.3 \micron^{-1}$:
\begin{equation*}
\begin{split}
a(x) = 1+0.17699y-0.50447y^2-0.02427y^3 +\\
+0.72085y^4+0.01979y^5-0.77530y^6+0.32999y^7
\end{split}
\end{equation*}
and
\begin{equation*}
\begin{split}
b(x) = 1.4138y+2.28305y^2+1.07233y^3-5.38434y^4 + \\
-0.62251y^5+5.30260y^6-2.09002y^7
\end{split}
\end{equation*}
where y=(x-1.82) and x=1/$\lambda$.\\

\subsection{Spectral templates of type Ia Supernovae} 

 We use the \citet{Hsiao2007ApJ...663.1187H} restframe SNe Ia spectra and light curve templates because they are most complete in temporal and wavelength coverage. The multi-epoch spectral templates range from -20 days before B$_{max}$ to 80 days after B$_{max}$, and were constructed by averaging a large number of observed spectra of nearby SN Ia between 0.1 and 2.5 $\mu$m. We tested the \citet{Nugent2002PASP..114..803N} Type Ia Branch-normal templates, and find that they produce similar results.

\section{Methods}
\label{methods}

In this section we describe our Monte Carlo simulation in which we randomly place SNe Ia spectrum templates in KINGFISH galaxies and calculate the dust affected spectrum. The main goal of our Monte Carlo simulation is to create probability plots of SNe Ia color excess as a function of galactocentric distance, and to place uncertainties on the color excess for different galaxy morphology types. We then compare our model to observed spectra of SNe Ia in the KINGFISH galaxies. Finally, we use the generated statistics in the Monte Carlo simulation to determine the absorption-to-reddening ratio $R_V$ of dust in host galaxies of a different sample of observed Ia SNe.

\subsection{Statistics with KINGFISH sample}

To develop color excess statistics we first investigate the relationship between the SNe dust extinction probability and position in the host galaxy. We also look for differences of the effect of dust extinction in galaxies with different morphological classifications. Our procedure was as follows.

\begin{figure*}[]
\begin{tabular}{ccc}
\hspace{-1.5cm}
\includegraphics[trim=25mm 5mm 30mm 5mm, width=7cm, clip=true]{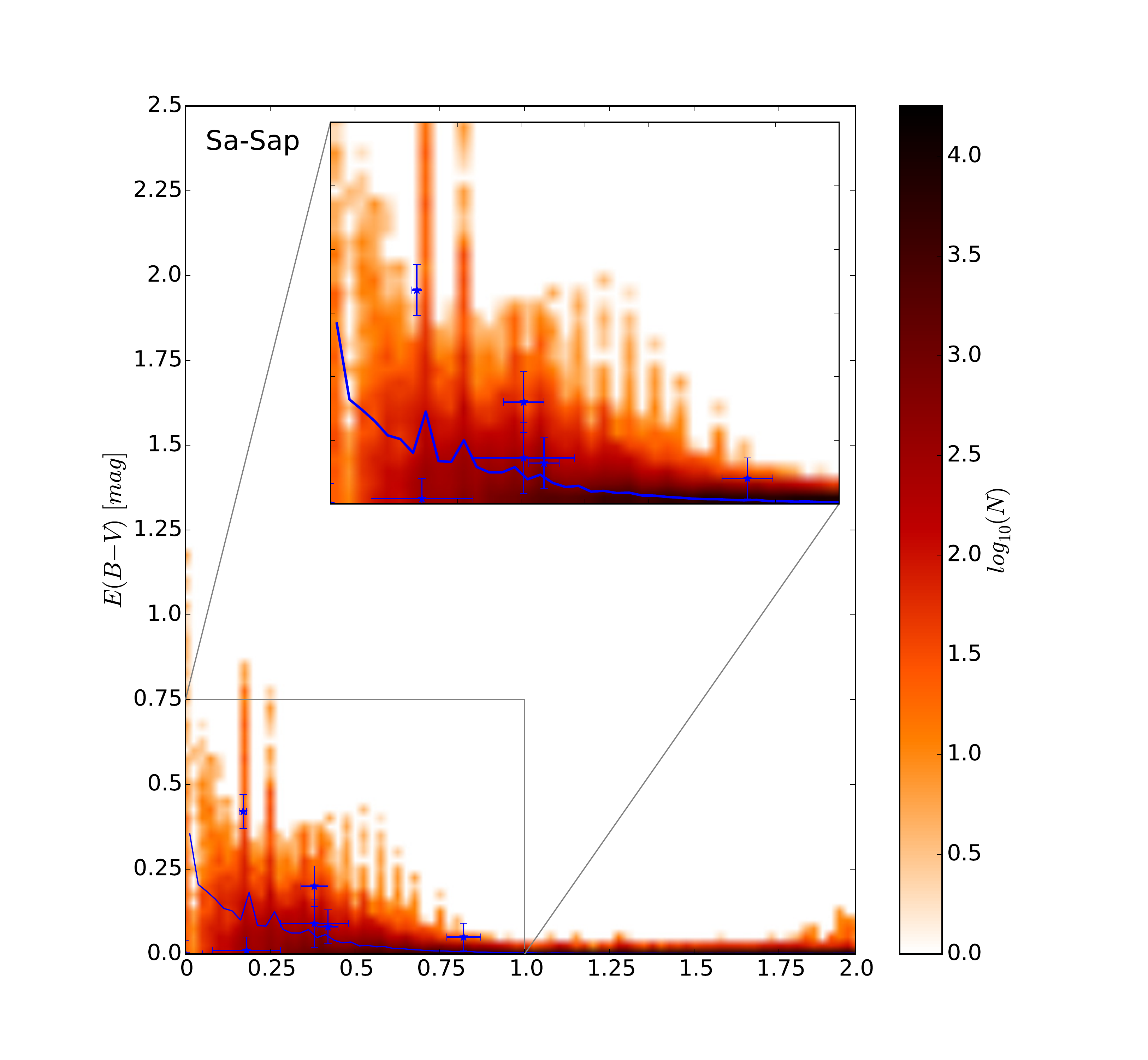} &
\hspace{-0.5cm}
\includegraphics[trim=25mm 5mm 30mm 5mm, width=7cm, clip=true]{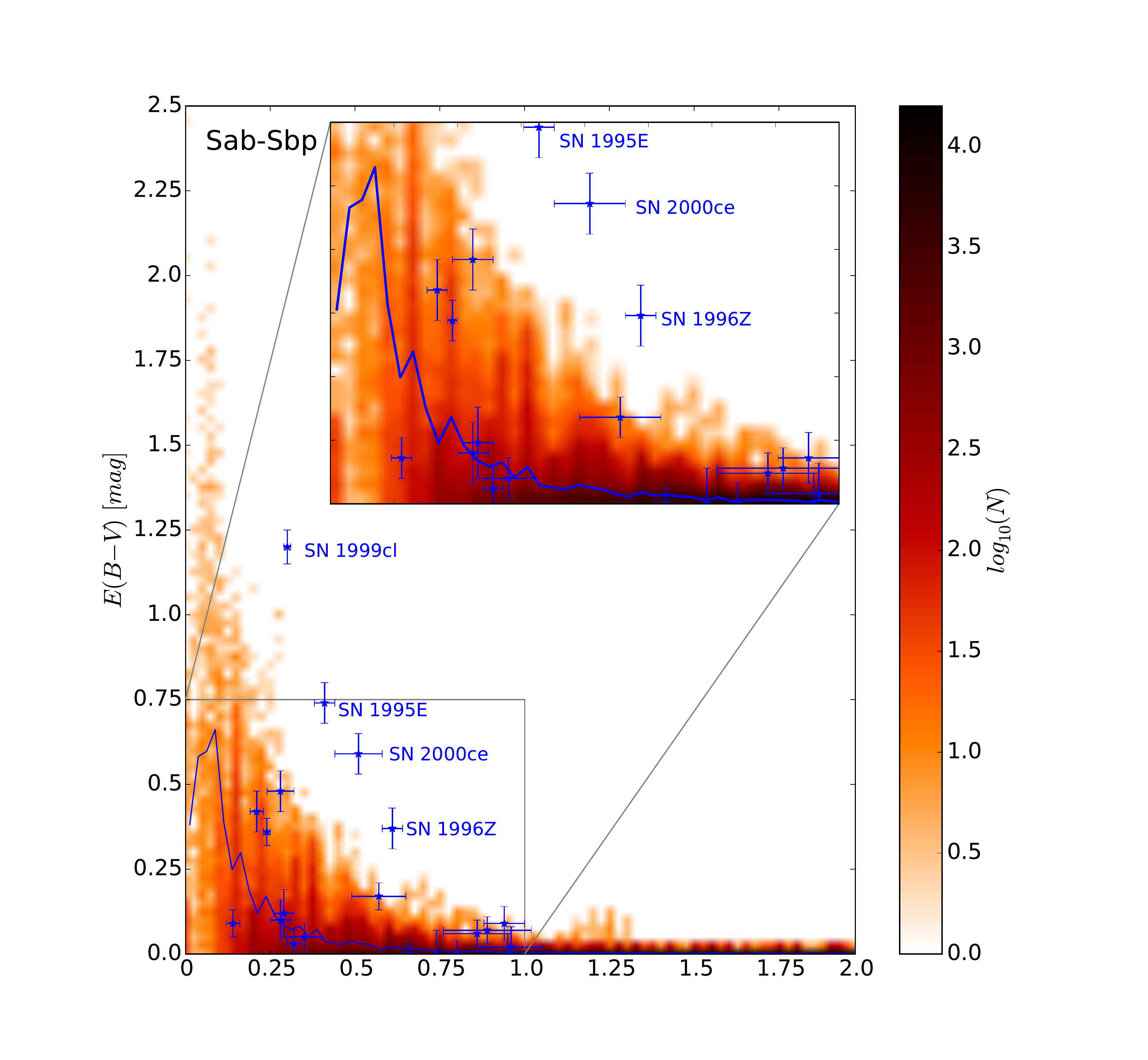} &
\hspace{-0.5cm}
\includegraphics[trim=25mm 5mm 30mm 5mm, width=7cm, clip=true]{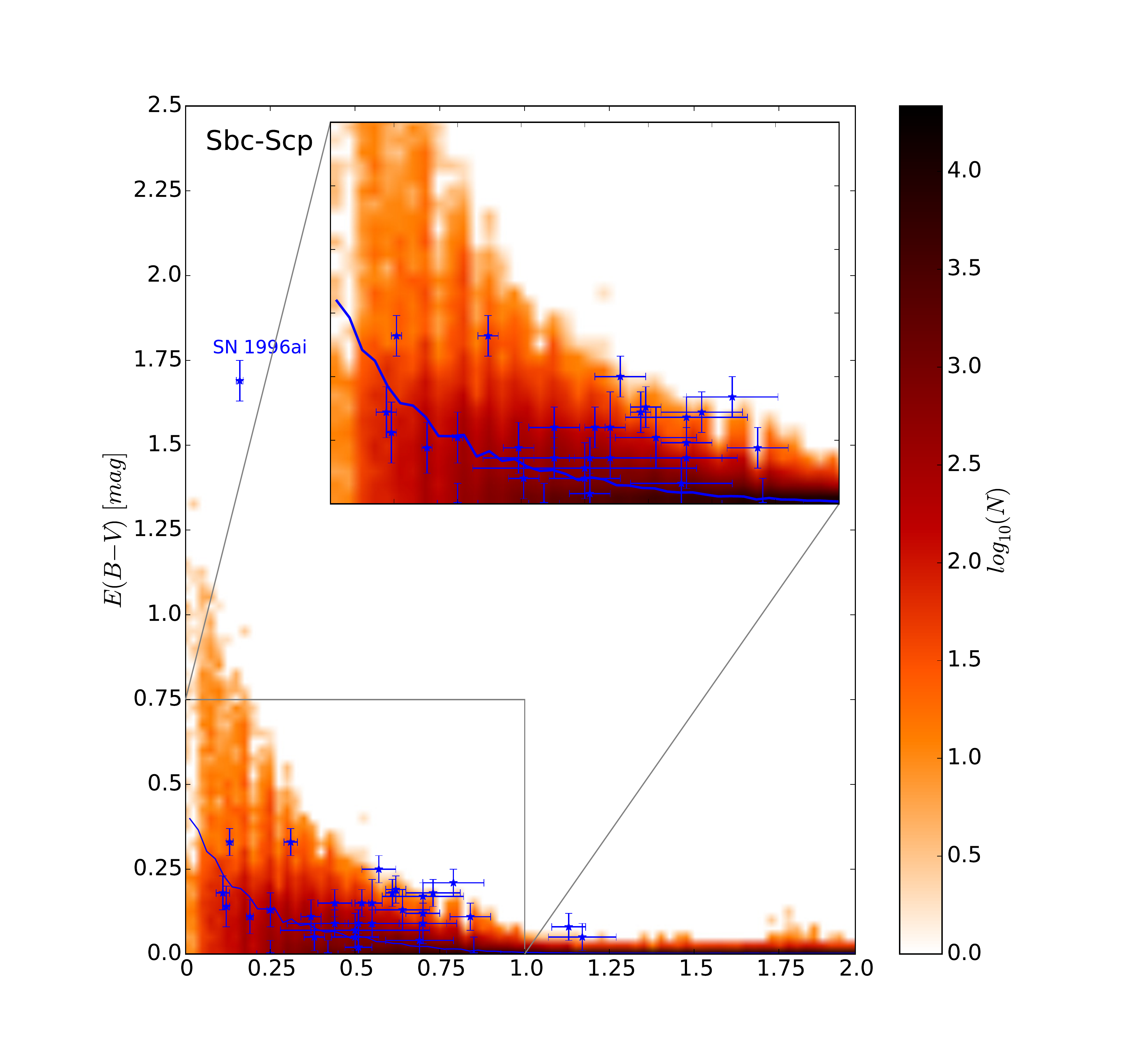} \\ 
\hspace{-1.5cm}
\includegraphics[trim=25mm 5mm 30mm 5mm, width=7cm, clip=true]{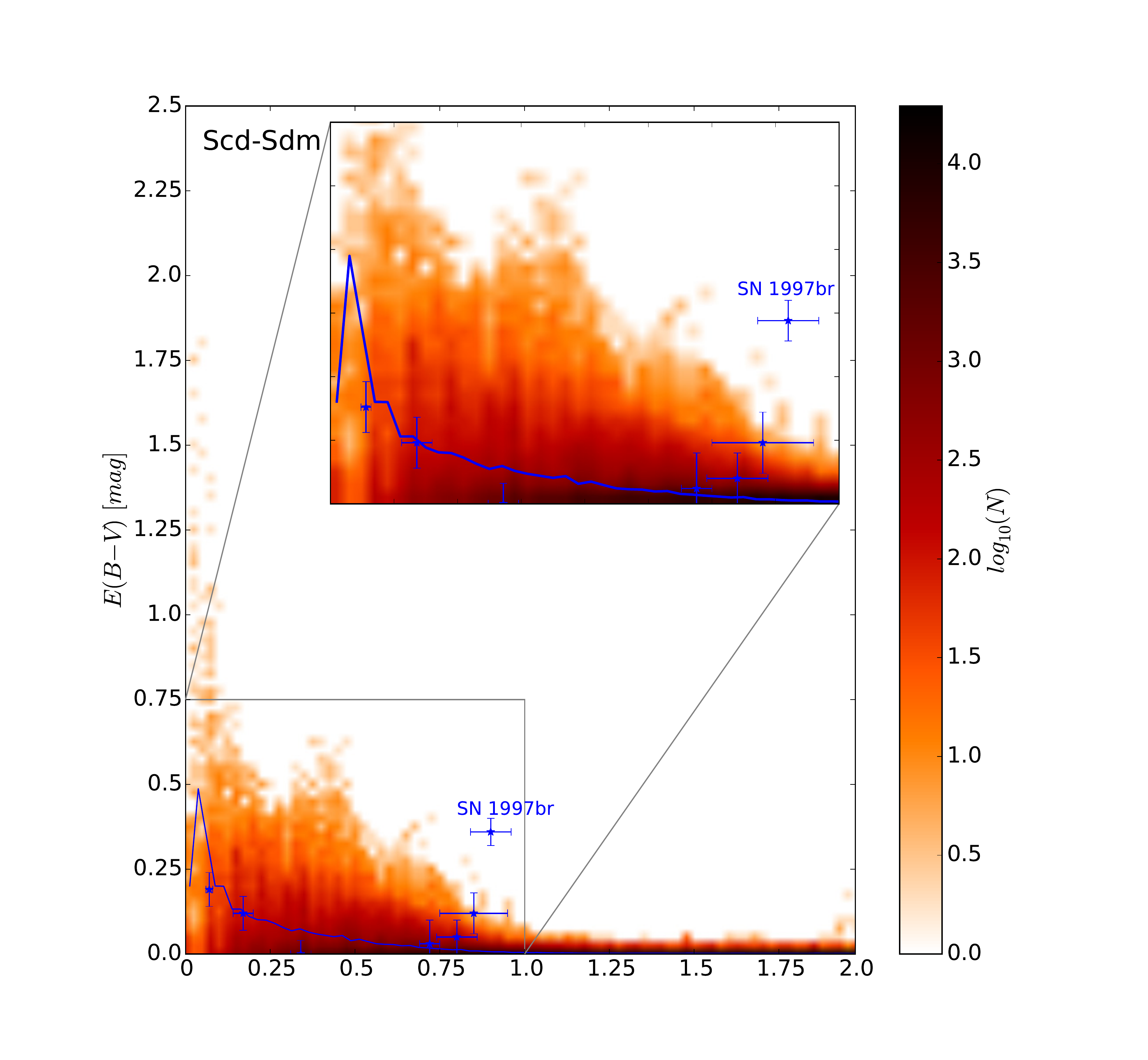} &
\hspace{-0.5cm}
\includegraphics[trim=25mm 5mm 30mm 5mm, width=7cm, clip=true]{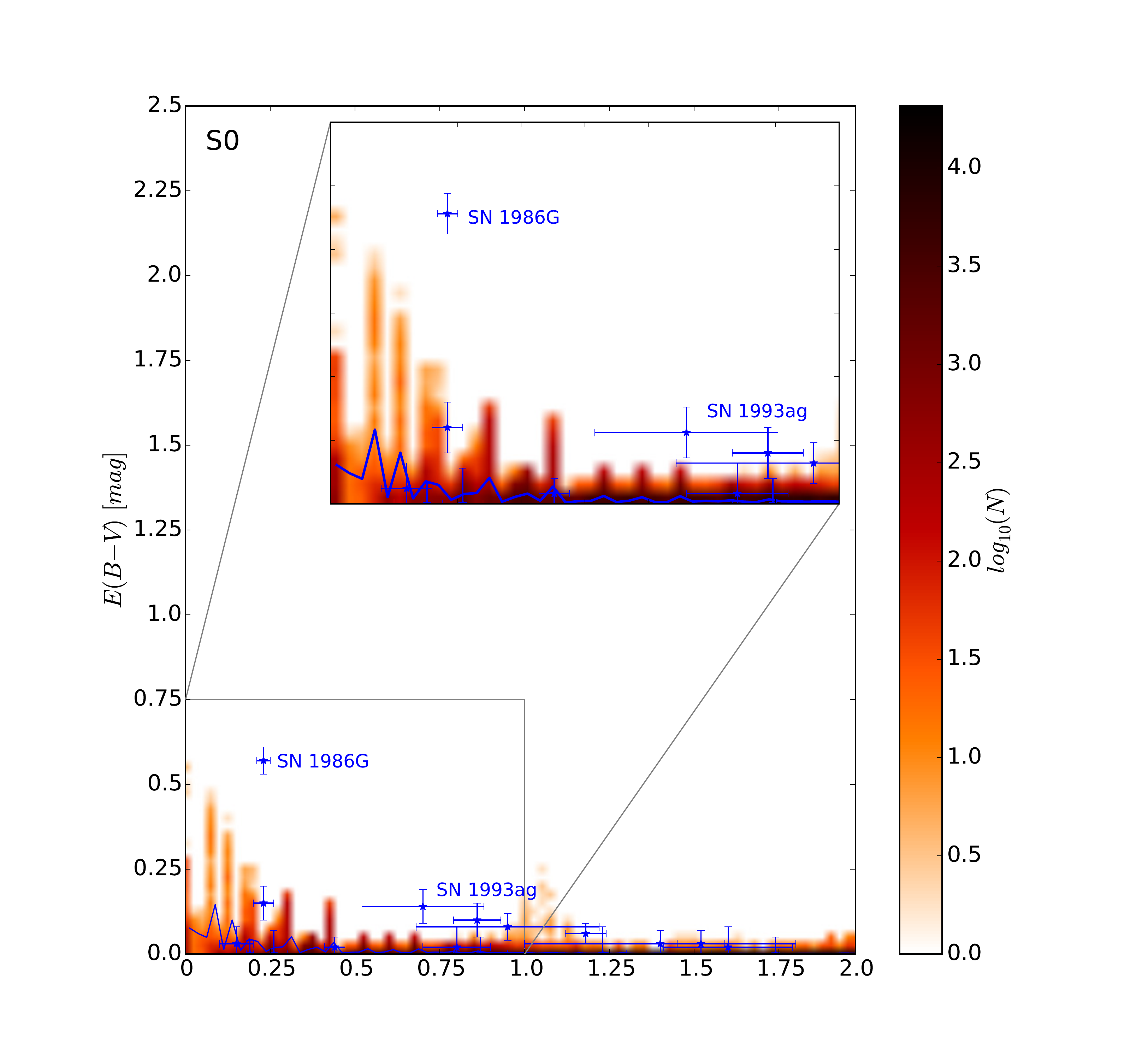} &
\hspace{-0.5cm}
 \includegraphics[trim=25mm 5mm 30mm 5mm, width=7cm, clip=true]{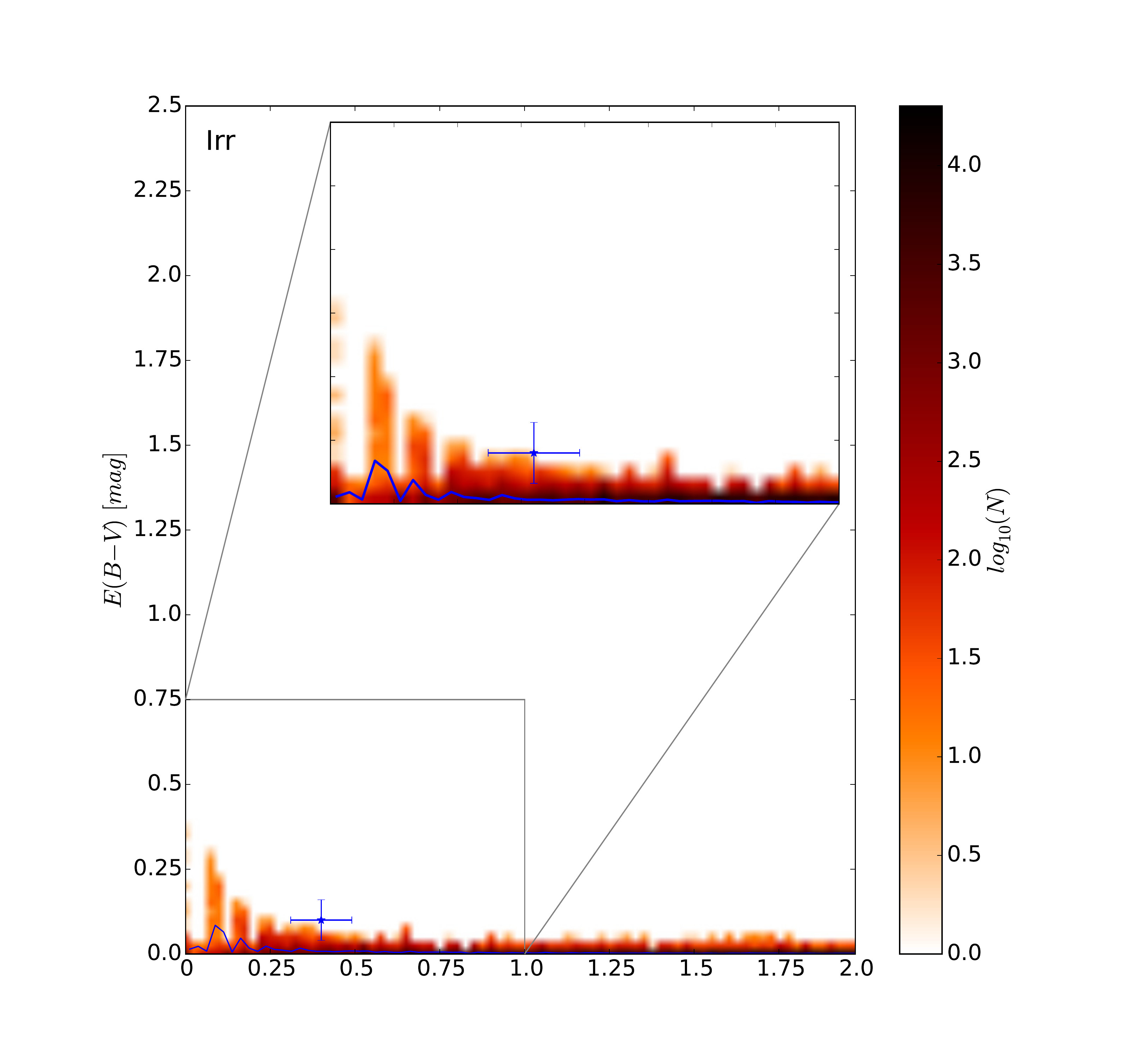} \\ 
\end{tabular} 
\caption{E(B-V) vs. normalized galactocentric distance probability plots for each of the galaxy groups: Sa-Sap, Sab-Sbp, Sbc-Scp, Scd-Sdm, S0 and Irregulars. The logarithmic color scale indicates the probability that a SN at a certain distance will have a certain color excess E(B-V), dark for high probability, and bright for low probability. The blue crosses indicate observed SNe Ia from the \citet{Wang2006ApJ...645..488W} sample. The blue lines, also shown in the insets and Figure~\ref{fitted}, are weighted means of E(B-V) for bins of size 0.025 $R/R_{25}$.}
\label{fig all_colorplots_3x2}
\end{figure*}

  The first step was to randomly place the peak brightness template spectrum 100000 times in each of our KINGFISH galaxies, within a radius of 2 $R_{25}$. We assume a peak absolute magnitude $M_B$ = -19.5 mag \citep{1999ApJ...522..802S}. The template supernovae are redshifted, and recalculated to observer's frame using the galaxy distances in Table~\ref{Kingfishgalaxies}.
The random parameters are projected (\textit{x,y}), centered on the galaxy nucleus, and \textit{ext}, a parametrization of the extinction,  uniformly randomized, corresponding to the fraction of the total amount of dust along the line of sight. A SN in front of the galaxy would have \textit{ext} = 0, one behind the galaxy \textit{ext} = 1.
For each instance, we calculate $A_V$ from equation (2) multiplied by \textit{ext}, and apply the CCM extinction law with $R_V$ = 3.1 to calculate $A_{\lambda}$/$A_V$, thus determining the selective extinction with $\lambda$, which, we then subtract from the SN Ia template.
In the second step we calculate the observed (B-V) colors, by convolving Bessel B and V filter passbands \citep{Bessell1990PASP..102.1181B} with the extincted SN spectrum.
The unreddened, intrinsic $(B-V)_{intrinsic}$ = -0.06 mag is calculated from the template spectrum with the same B and V passbands ($A_V$ = 0 mag). 
The outputs are $(B-V)_{observed}$, the color excess $E(B-V) = (B-V)_{observed} - (B-V)_{intrinsic}$, and the projected galactocentric distance in pixels. In Figure~\ref{fig all_colorplots_3x2} we show the distribution of E(B-V) probability with projected $R/R_{25}$ for each galaxy group. E(B-V) peaks near the center, and decreases with radius, as expected, although peak E(B-V) is highest for Sab-Sb types and lowest for S0 and Irregulars.

\subsection{Comparison with observed SNe Ia in KINGFISH galaxies}

There are 15 SNe Ia that occurred in KINGFISH galaxies (Table~\ref{AllIaSNeinKINGFISH}). We found spectra for five SNe in the online Supernova Spectrum Archive (SUSPECT)\footnote{\url{http://www.nhn.ou.edu/~suspect/}}: SN 1981B in NGC 4536, SN 1989B in NGC 3627, SN 2002bo in NGC 3190, SN 2006X in NGC 4321 and SN 1966J in NGC 3198. The spectrum of SN 1966J is digitized from a photographic plate, obtained with the prismatic nebular spectrograph on the Crossley reflector at the Lick Observatory. It is valuable for classification purposes, but as the flux is not properly calibrated \citep{2000PASP..112.1433C} it is not useful for our purpose. The units of the SN 1981B spectra are unknown, and do not agree with commonly used units by Branch (e.g. in \citet{1983ApJ...270..123B, 1981ApJ...244..780B}). General information about SN 1989B, SN2002bo and 2006X are summarized in Table~\ref{SNe in KINGFISH}. \\

Having calculated the dust mass in each galaxy pixel, we compare the simulation results to the observed spectra of these historical SNe to see if we can reproduce the observed quantities for $R_V$ = 3.1.

\begin{figure}
\begin{center}
\includegraphics[trim=7mm 0mm 7mm 0mm, width=9cm]{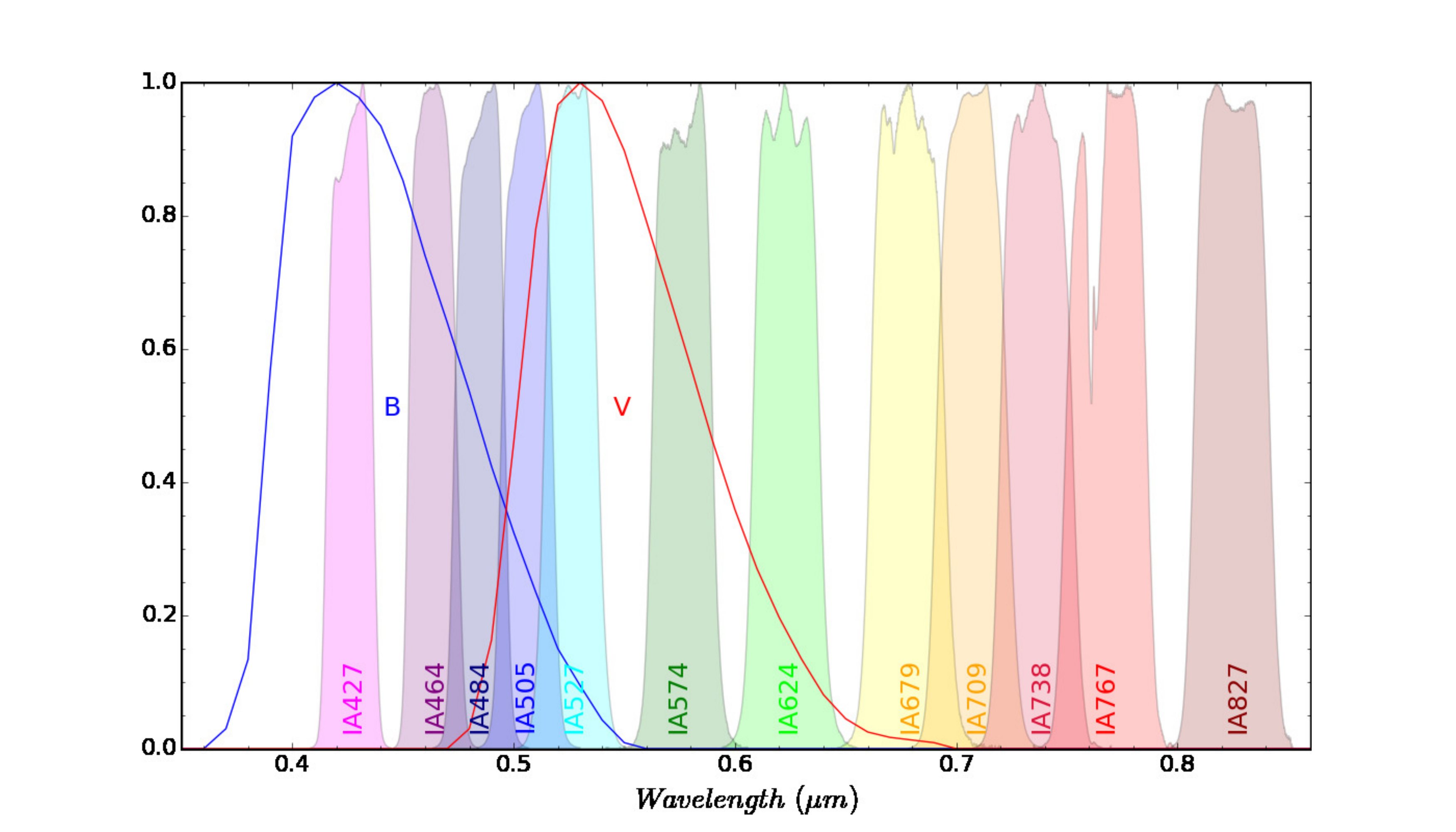}
\vspace{-5mm}
\caption{Subaru Telescope Intermediate Band filters (shaded), and Bessel B (blue line) and V (red line) filters.}
\label{figpassbands}
\end{center}
\end{figure}

The procedure of the comparison is as follows:

\begin{enumerate}

\item Match the RA and Dec of the observed SN Ia to the (X,Y) coordinates of the KINGFISH map of the host galaxy.
\item Calculate the extinction curves using $R_V$ = 3.1 for different $A_V$, from 0 to 3.5 in 0.1 steps
\item $A_V$ in the preceeding step are applied to the corresponding restframe template spectra, generating 35 reddened spectra for each epoch.
\item Each spectrum is redshifted into the observer's frame and the appropriate Milky Way extinction applied. 
\item The observed and redshifted templates are convolved with the Subaru intermediate filter passbands (Figure~\ref{figpassbands}, \citet{2004sgyu.conf..107T}) in order to have a number of photometric comparison points.
\item To determine $A_V$ that best matches the observed spectra we apply the least square test, 
\begin{equation}
\label{leastsqtesteq}
\mathrm{\chi^2=R^2=\sum[OBS_i-SIM_i]^2} ,
\end{equation} 
where $OBS_i$ and $SIM_i$ are the observed and simulated photometry in the Subaru filters. 

\item To calculate E(B-V), the best matching template spectra are convolved with Bessel B and V filter passbands (\citet{Bessell1990PASP..102.1181B}, Figure~\ref{figpassbands}) and using $(B-V)_{instrinsic}$ calculated for each epoch.
\item We compare the best fit $A_V$, to the calculated $A_V$ from the dust assuming ext=1.

\item We also calculate a set of model spectra using different values of $R_V$ (from 0.5 to 5.5 in 0.1 steps), which we then compare to the observed spectra as described in steps 2-7. The results are shown in Table~\ref{table KINGFISH bestfit}.

\end{enumerate}

\begin{table}
\caption{Overview of Ia SNe in the KINGFISH sample}
\vspace{-5mm}
\label{AllIaSNeinKINGFISH}
\begin{center}
\begin{tabular}{lllll }
\hline\hline
SN    &   Host galaxy  & Mag.  &  Type &  Spectra \\
\hline\hline
SN 1957A  & NGC 2841   &  14.0  &  Ia-p &    \\
SN 1966J  & NGC 3198   &  13.0  &  Ia   &  SUSPECT \\
SN 1980N  & NGC 1316   &  12.5  &  Ia   &	 \\
SN 1981B  & NGC 4536   &  12.3  &  Ia   &  SUSPECT  \\
SN 1989B  & NGC 3627   &  13.0  &  Ia   &  SUSPECT  \\
SN 1989M  & NGC 4579   &  12.2  &  Ia   &	 \\
SN 1994ae & NGC 3370   &  15.4  &  Ia   &	 \\
SN 1999by & NGC 2841   &  15.0  &  Ia-p &	 \\
SN 2002bo & NGC 3190   &  15.5  &  Ia   &  SUSPECT \\
SN 2002cv & NGC 3190   &  19.0  &  Ia   &	 \\
SN 2006dd & NGC 1316   &  15.0  &  Ia   &	 \\
SN 2006mr & NGC 1316   &  15.6  &  Ia   &	 \\
SN 2006X  & NGC 4321   &  17.0  &  Ia   &  SUSPECT \\
SN 2007on & NGC 1404   &  14.9  &  Ia   &	 \\
SN 2011iv & NGC 1404   &  12.8  &  Ia   &	 \\
\hline\hline
\end{tabular}
\end{center}
\vspace{-3mm}
\textbf{Notes.} Magnitudes and type were taken from the IAU Central Bureau for Astronomical Telegrams (CBAT) (\url{http://www.cbat.eps.harvard.edu/lists/Supernovae.html}).
\end{table}

\begin{table*}
\caption{General information on SN 1989B, SN2002bo and 2006X}
\vspace{-5mm}
\label{SNe in KINGFISH}
\begin{center}
\begin{tabular}{ l l l l l l l l }
\hline\hline
\sc Supernova  & \sc Galaxy & \sc Distance & $M_B$(max) & \sc $B_{mag}$ & $E(B-V)_{Gal}$    & \sc Spectra epoch          & \sc Spectrum \\
               &            & $(Mpc)$      & $(mag)$    &  $(mag)$      &  (mag)            & \sc relative to B$_{max}$  &  \sc references  \\
\hline\hline
SN 1989B  & NGC 3627 & 11.07\footnotemark[1]  & -19.48\footnotemark[1] & 12.34(0.05) & 0.032  & 0, 6, 11, 21, 22, 31, 52 &  {\citet{Barbon1990A&A...237...79B}} \\ 
SN 2002bo & NGC 3190 & 21.6\footnotemark[2]  & -19.41\footnotemark[2] & 14.06(0.06) & 0.027 & -4, -3, -1, 4, 28      & {\citet{Benetti2004MNRAS.348..261B}} \\ 
SN 2006X  & NGC 4321 & 15.2\footnotemark[3]   &  -19.1\footnotemark[4]  & 15.41(0.01) & 0.026 &-6, 0, 2, 6, 8, 12, 13   &{\citet{2009PASJ...61..713Y}}   \\ 
\hline\hline
\end{tabular}
\end{center}
\vspace{-3mm}
\footnotetext[1]{determined using Cepheid data \citep{1999ApJ...522..802S}}
\footnotetext[2]{derived from redshift in \citet{Benetti2004MNRAS.348..261B}}
\footnotetext[3]{determined using Cepheid data \citep{WangSN2006X2008ApJ...677.1060W}} 
\footnotetext[4]{determined using Cepheid data \citep{Wang2008ApJ...675..626W}}
\end{table*}

\subsubsection{SN 1989B in NGC 3627}
\begin{figure}
\begin{center}
\includegraphics[width=9cm]{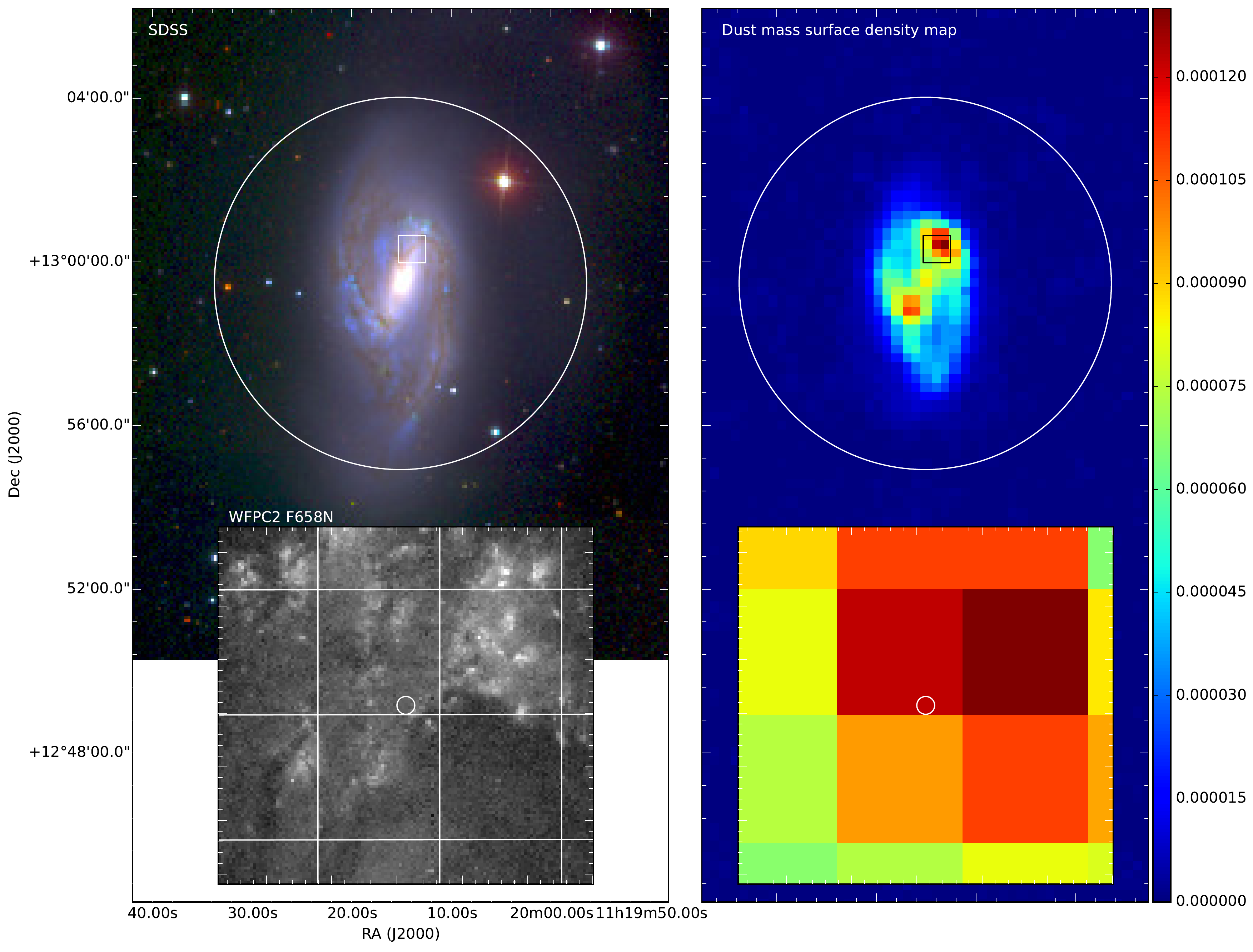}
\vspace{-5mm}
\caption{Top left: Shown is a SDSS gri color image of NGC 3627. The radius of the white circle corresponds to the de Vaucouleur radius $R_{25}$. Bottom left: A \textit{Hubble Space Telescope} WFPC2/F658N image of the region inside the white box. The small white circle denotes the position of SN 1989B. The radius of 1'' corresponds to the position uncertainty. The gird corresponds to pixels of the Herschel's 500 $\mu$m map. Top right: Dust mass surface density map of NGC 3627. The region inside the black box is shown enlarged bottom right. The white circle denotes the position of the SN, in a high dust density region.}
\label{figHSTSN1989B}
\end{center}
\end{figure}

SN 1989B is a normal SNe Ia which occurred in a bright spiral arm of NGC 3627, 15''W and 50''N of its nucleus. Figure~\ref{figHSTSN1989B} shows the position of the SN in the dust mass surface density map, compared to optical SDSS and \textit{Hubble Space Telescope} images. \citet{Wells1994AJ....108.2233W} observed maximum light in B at a magnitude of B = 12.34 $\pm$ 0.05 mag and derived a color excess of E(B-V) = 0.37 $\pm$ 0.03 mag from UBVRIJHK photometry, using Galactic reddening $E(B-V)_{Gal}$ = 0.032 mag, and assuming $R_V$ = 3.1 extinction. We simulated the spectra for days 0, 6, 11, 21, 22, 32 and 52, assuming a distance of 11.07 Mpc and an absolute brightness of $M_B$(max)=-19.48 mag \citep{1999ApJ...522..802S}. The total dust extinction, calculated from the KINGFISH dust mass column density map of NGC 3627, in V band at the position of SN 1989B is $A_V$ = 3.0 mag, in case of \textit{ext} = 1, when all of dust is taken into account ($\sigma_{dust}$ = 9.667 $\times$10$^{-5} g$  $cm^{-2}$).

\begin{figure}
\begin{center}
\includegraphics[trim=7mm 0mm 7mm 0mm, width=9cm]{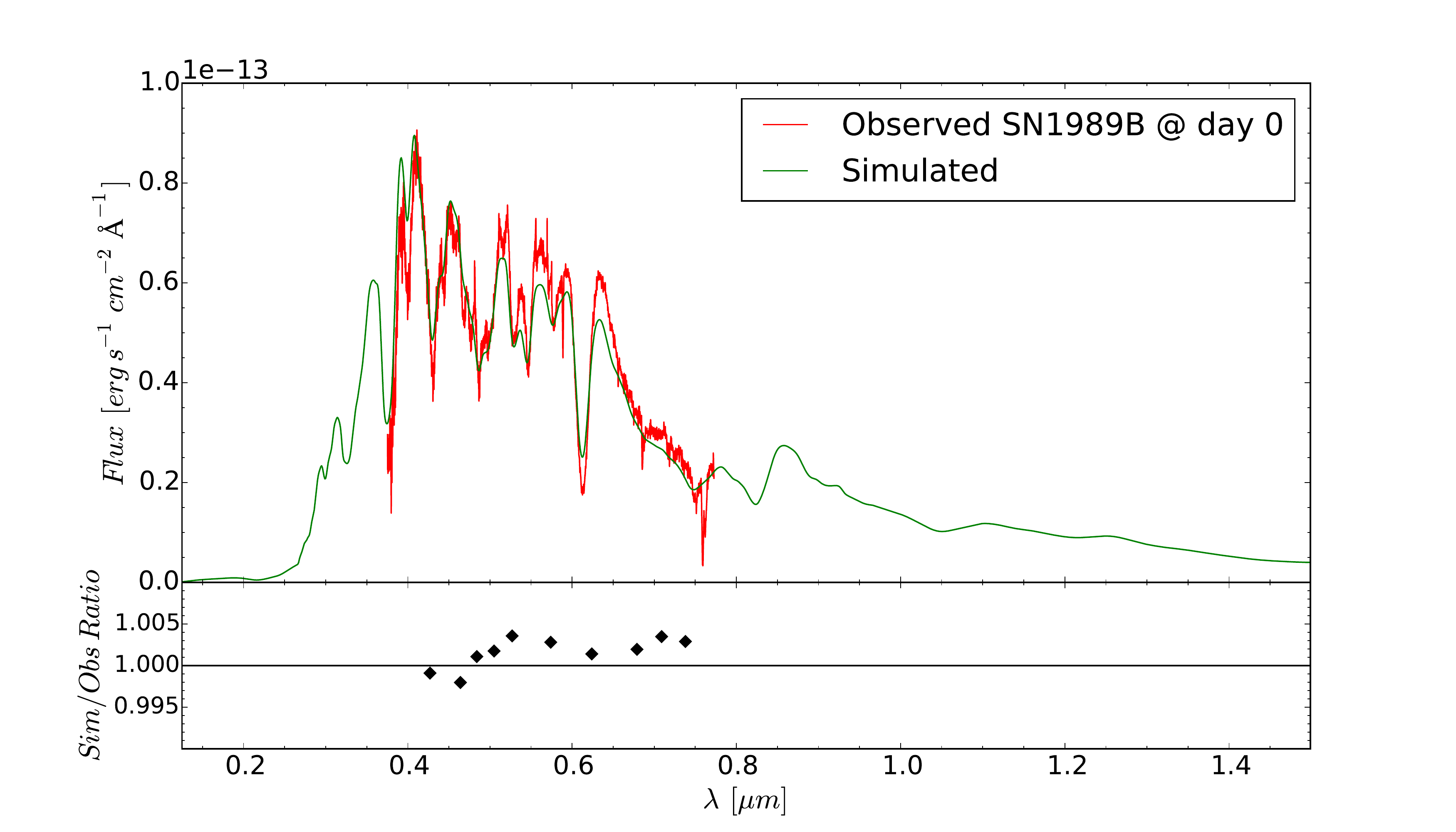}
\vspace{-5mm}
\caption{Comparison between the observed SN 1989B spectrum at peak brightness and the spectral template, for $R_V$ = 3.1. The template best matches the observed spectrum after applying CCM extinction law with $A_V$ = 1.2 mag. The apparent brightness at maximum light of the simulated best fit spectrum is $m_B$ = 12.46 mag, and E(B-V) = 0.37 mag, which is consistent with the observations. The bottom plot shows the ratio between simulated and observed fluxes calculated by convolving the spectra with Subaru Telescope Intermediate Band filters.}
\label{fig SN 1989B}
\end{center}
\end{figure}

We derive from the simulated best fit spectra E(B-V) = 0.27 $\pm$ 0.07 mag and $A_V$ = 0.95 $\pm$ 0.19 mag in case of applying an extinction law with $R_V$ = 3.1. Figure~\ref{fig SN 1989B} shows the best fitted  simulated spectrum to the observed spectra \citep{Barbon1990A&A...237...79B} at peak brightness. 
The results are consistent with the \citet{Wells1994AJ....108.2233W} observations. The amount of dust at the location where SN 1989B occurred is enough to cause the observed reddening.

\begin{figure}
\begin{center}
\includegraphics[trim=7mm 0mm 7mm 0mm, width=9cm]{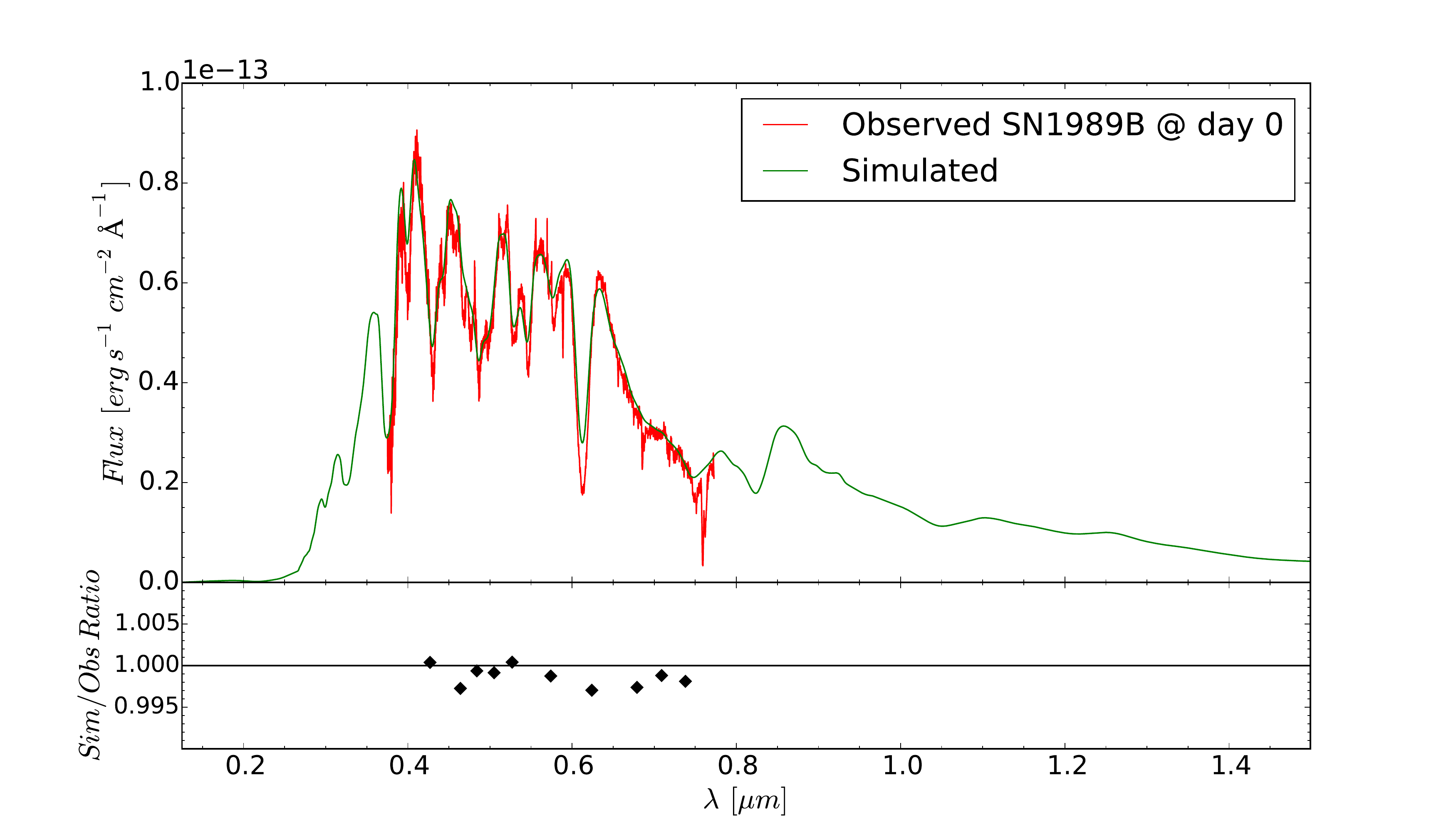}
\vspace{-5mm}
\caption{Comparison between the observed SN 1989B spectrum at peak brightness and the spectral template. The best fit spectrum is obtained by applying the extinction law with $R_V$ = 2.2 and $A_V$ = 1.1 mag. The apparent brightness at maximum light of the simulated best fit spectrum is $m_B$ = 12.48 mag, and E(B-V) = 0.48 mag. The bottom plot shows ratios between simulated and observed fluxes calculated by convolving the spectra with Subaru Telescope Intermediate Band filters.}
\label{fig SN 1989B besfit}
\end{center}
\end{figure}

We vary the $R_V$ value, and apply the extinction law until the simulated spectrum best fits the observed ones. We derive an average absorption-to-reddening ratio $R_V$ = 3.16 $\pm$ 1.54 mag and E(B-V) = 0.34 $\pm$ 0.21 mag from observed spectra at seven different epochs. The results are given in Table~\ref{table KINGFISH bestfit}, and Figure~\ref{fig SN 1989B besfit} shows of the best fit simulated spectrum to the observed spectra \citep{Barbon1990A&A...237...79B} at peak brightness.

\subsubsection{SN 2002bo in NGC 3190}
SN 2002bo is a typical Branch normal SN Ia at visible and infrared wavelengths. \citet{Benetti2004MNRAS.348..261B} determined its peak absolute magnitude, $M_B$(max) = -19.41 mag, E(B-V) $\sim$ 0.47 mag and a distance of 21.6 Mpc. Milky Way extinction along the line of sight is E(B-V)$_{Gal}$ = 0.027 mag \citep{Schlegel1998ApJ...500..525S}. 
The dust mass column density in the KINGFISH dust maps at the position of SN 2002bo, for \textit{ext} = 1, is $\sigma_{dust}$=(3.155$\pm$0.112)$\times10^{-5} g$ $cm^{-2}$, corresponding to $A_V$ = 0.98 $\pm$ 0.04 mag.

\begin{figure}
\begin{center}
\includegraphics[trim=25mm 5mm 25mm 20mm, width=9cm, clip=true]{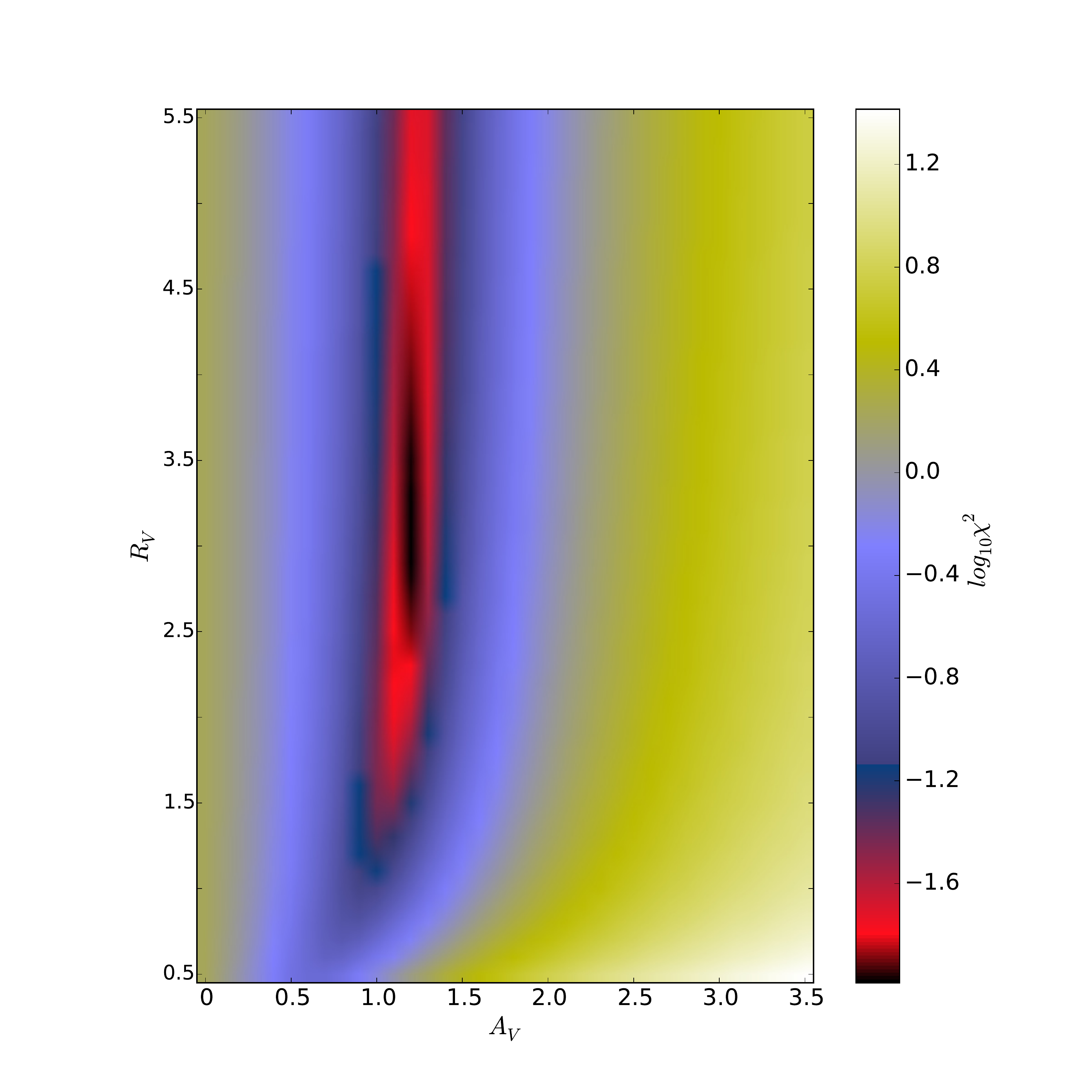}
\vspace{-5mm}
\caption{The distribution of $\chi^2$ values for the matching procedure for SN 2002bo, 4 days before peak brightness. We apply an extinction law with a range of different $R_V$ and $A_V$ values to the spectrum template, and calculate $\chi^2$ using observed and simulated Subaru intermediate passbands photometry.
In this case, the template best matches the observed spectrum after applying the extinction law with $R_V$ = 3.1 and $A_V$ = 1.2 mag. However, the difference in $\chi^2$ for $R_V$ $\sim$ 2.7 - 3.3, at $A_V$ = 1.2 mag is very small.}
\label{fig:leastsqaremap}
\end{center}
\end{figure}

\begin{figure}
\begin{center}
\includegraphics[trim=7mm 0mm 7mm 0mm, width=9cm]{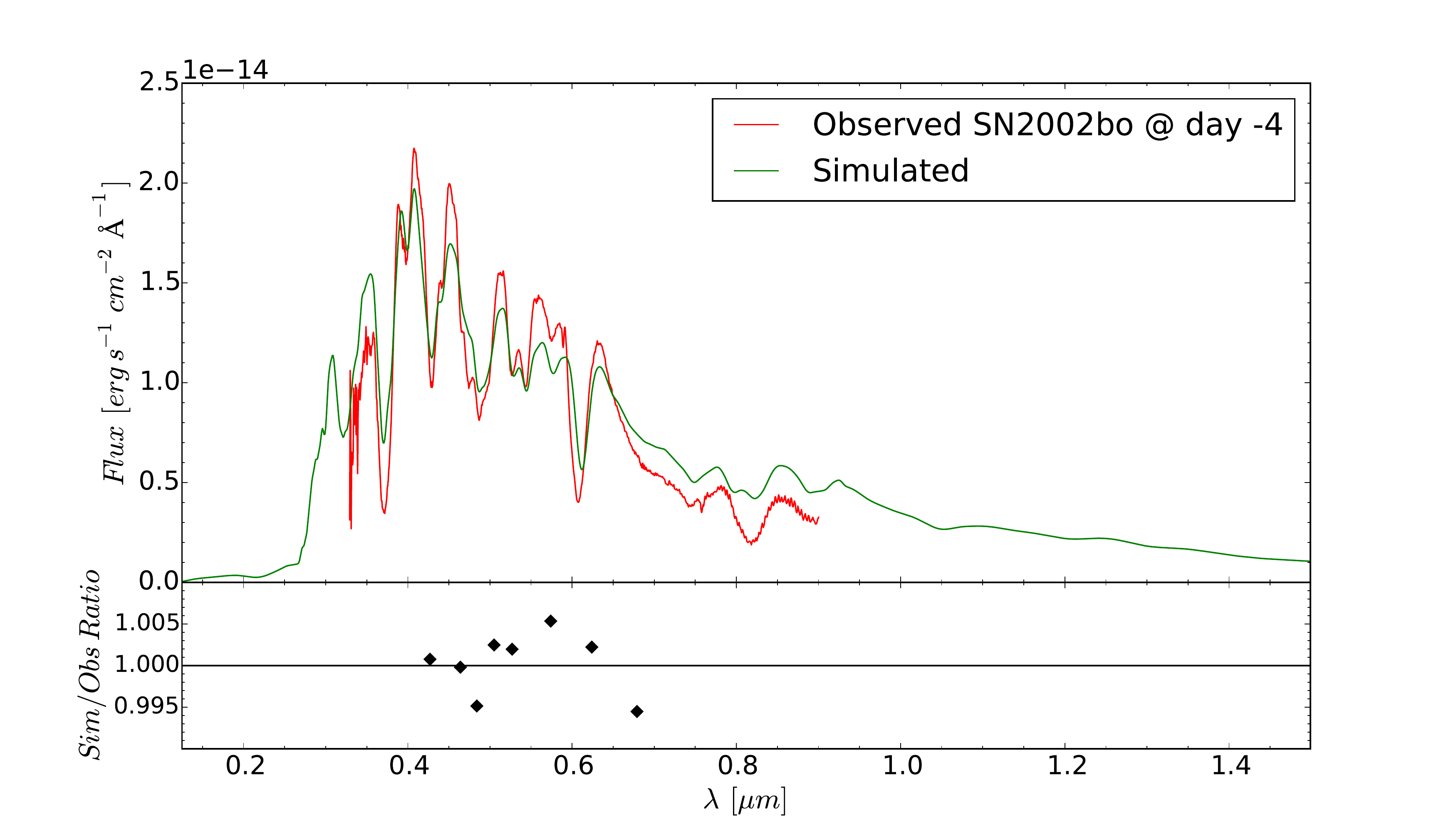}
\vspace{-5mm}
\caption{Comparison between the observed SN 2002bo spectrum 4 days before maximum brightness and the spectral template. The best fit spectrum is obtained by applying extinction law with $R_V$ = 3.1 and $A_V$ = 1.2 mag. The bottom plot shows ratios between simulated and observed fluxes calculated by convolving the spectra with Subaru Telescope Intermediate Band filters.}
\label{fig SN_2002bo}
\end{center}
\end{figure}

\begin{figure}
\begin{center}
\includegraphics[width=9cm]{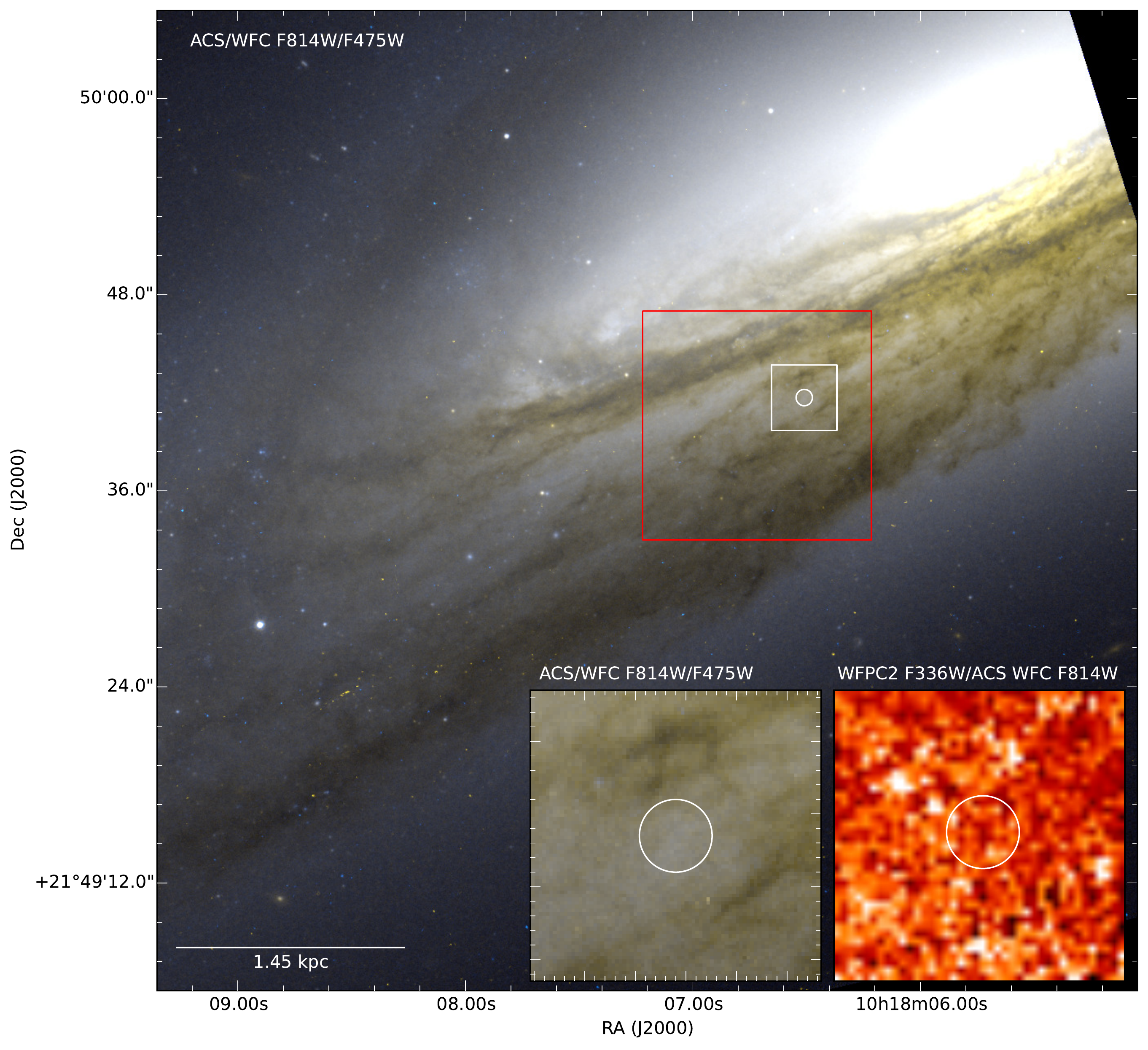}
\vspace{-5mm}
\caption{\textit{Hubble Space Telescope} ACS/WFC F814W/F475W color image of NGC 3190. The position of SN 2002bo is marked with the small, white circle.  The large red square (14'' x 14'') is the SPIRE map pixel. The region within the small white box is shown enlarged in the two inset figures, where the r = 0.5'' circle corresponds to the position uncertainty of the SN. Bottom left: Zoomed in view of the ACS image. Bottom right: WFPC2 F336W - ACS/WFC F814W color map. In both inset images, the clumpy distribution of dust is noticeable.}
\label{ngc3190}
\end{center}
\end{figure}

For $R_V$ = 3.1, we find that the best fitting template to the observed spectra have $R_V$ = 1.34 $\pm$ 0.19 mag and E(B-V) = 0.41 $\pm$ 0.07 mag. This $A_V$ value is significantly larger ($\sim$ 35$\%$) than the value calculated from the Herschel derived dust masses. Possible reasons for this discrepancy are:

   a) {\it Dust mass uncertainty}. Total dust masses for this galaxy range from $10^{6.89}M_{\odot}$ (determined by Skibba) to $10^{7.19}M_{\odot}$ (determined by Draine). Our total integrated mass of $10^{6.97}M_{\odot}$ is $\sim$66$\%$ lower than the \citet{Draine2007ApJ...663..866D} mass, and $\sim$38$\%$ lower than the \citet{Gordon2008} mass, which is sufficient to reach the observed $A_V$.
   
   b) {\it $R_V \neq$ 3.1}. We recalculated the template spectra using values of $R_V$ = 0.5 - 5.5 in 0.1 steps, and varying $A_V$ between 0 and 3.5 in 0.1 steps for each value of $R_V$. We find the best fits to E(B-V) = 0.41 $\pm$ 0.07 mag and $R_V$ = 3.24 $\pm$ 0.97, which are consistent with the assumed $R_V$ = 3.1. Figure~\ref{fig:leastsqaremap} shows a least square colormap for best fit determination of the template to the observed spectra 4 days before peak brightness. Figure~\ref{fig SN_2002bo} shows the observed \citet{Benetti2004MNRAS.348..261B} spectra and our best fit computed spectrum for day -1 relative to maximum brightness.

 c) {\it Inhomogeneously distributed dust or clumpy dust.} At the distance of NGC 3190, a Herschel 500 $\mu$m pixel corresponds to $\sim$1.45 kpc. Figure~\ref{ngc3190} shows a \textit{Hubble Space Telescope} ACS/WFC (PropID: 10594) composite image of NGC 3190. SN 2002bo's position and the 500 $\mu$m SPIRE map pixel are marked. SN 2002bo is inside a spiral arm of NGC 3190 surrounded by clumpy dust features (ACS image); the dust is not smoothly distributed within the SPIRE pixel. If SN 2002bo lies within or behind a dense dust clump, this may explain the larger observed extinction compared to the average $A_V$ value calculated for the 500 $\mu$m pixel. The ratio of WFPC2/F336W (PropID: 11966) and ACS-WFC/F814W also shows that the dust is not smoothly distributed near SN 2002bo.

\subsubsection{SN 2006X in NGC 4321}
According to \citet{WangSN2006X2008ApJ...677.1060W}, SN 2006X is highly reddened by abnormal dust, and compared to other SNe Ia has a peculiar intrinsic color evolution and the highest expansion velocity ever published for a SN Ia. It seems to have a color excess of E(B-V) $\sim$ 1.42 mag with $R_V$ $\sim$ 1.48, much smaller than the Milky Way value of $R_V$ = 3.1, indicating that the dust surrounding SN 2006X has much smaller grain size than typical interstellar dust. They also note that most of the highly reddened SNe Ia with $E(B-V)_{Host}$ $>$ 0.5, tend to have $R_V$ values smaller than 3.1, which suggests that dust surrounding some highly extinguished SNe may be different from that observed in the Galaxy and used in our model. 
\citet{2009A&A...508..229P} observed that this SN has high degree of interstellar polarization, peaking in the blue part of the spectrum. Based on \citet{1975ApJ...196..261S}, \citet{2009A&A...508..229P} conclude that the dust mixture must be significantly different from typical Milky Way dust, with a lower total to selective extinction ratio $R_V$ than commonly measured in the Milky Way. They suggest that the bulk of the extinction is due to a molecular cloud.

\begin{figure}
\begin{center}
\includegraphics[trim=7mm 0mm 7mm 0mm, width=9cm, clip=true]{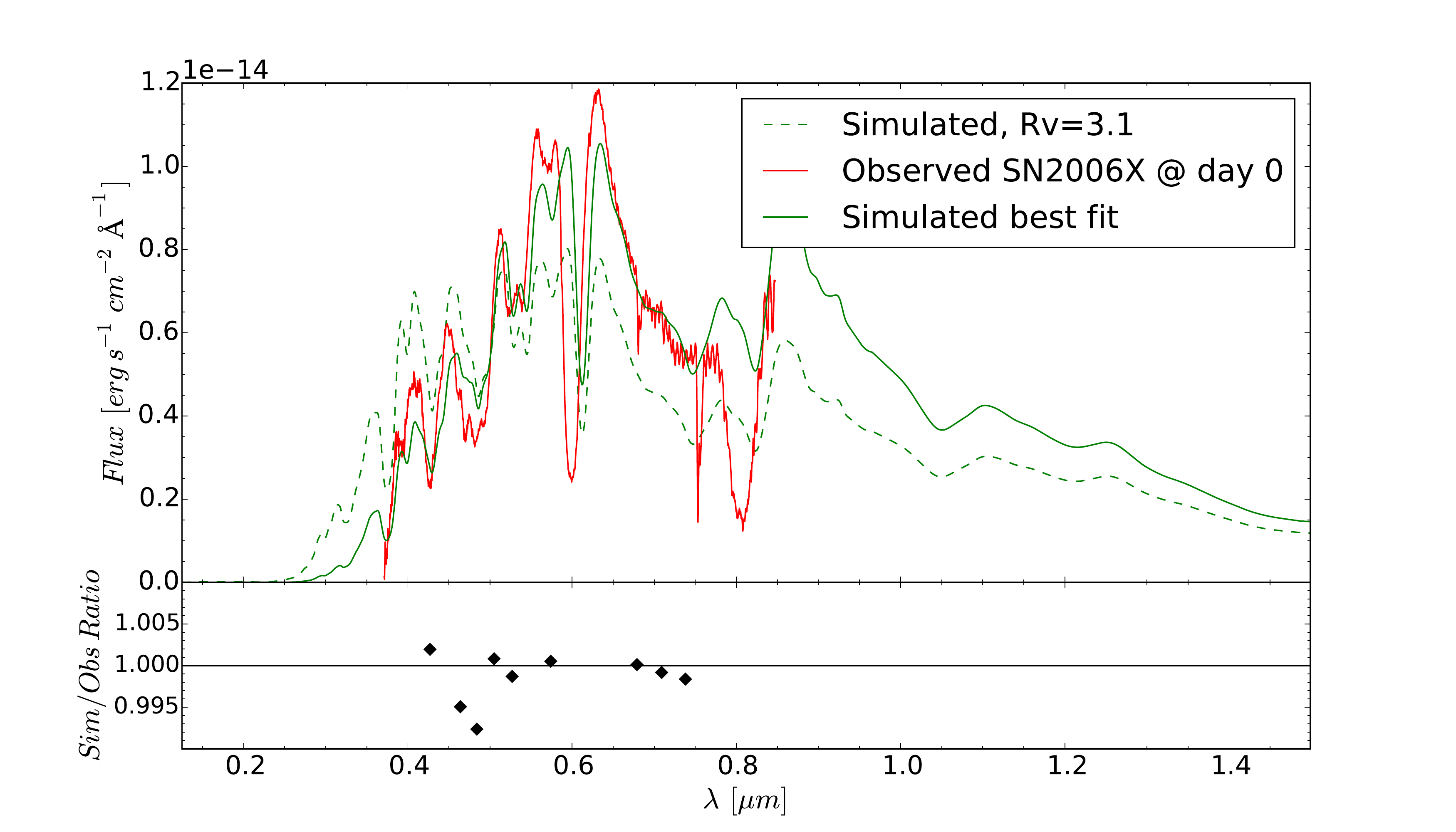}
\vspace{-5mm}
\caption{Comparison of template spectra and the observed spectrum of SN 2006X at peak brightness. The dashed line is the best fit template with $R_V$ = 3.1, and $A_V = 2.4$ mag. The solid green line is the best fit for an extinction law with $R_V$ = 1.6 and $A_V$ = 2.2 mag.} 
\label{fig SN 2006X}
\end{center}
\end{figure}

We used the Cepheid distance to NGC 4321 of 15.2 Mpc, and an absolute brightness at peak of $M_B$ = -19.1 mag \citep{Wang2008ApJ...675..626W} to best fit the simulated spectra to observations. 
At the position of SN 2006X in the extinction map of NGC 4321, $A_V$ = 1.05 mag (for $R_V$= 3.1), and,  Milky Way reddening along the line of sight is $E(B-V)_{Gal}$ = 0.026 mag. We determined from spectra at seven different epochs (Table~\ref{table KINGFISH bestfit}) an average $R_V$ = 1.46 $\pm$ 0.29, and E(B-V) = 1.39 $\pm$ 0.05 mag. This $R_V$ ratio is consistent with the values determined by others. The observed spectra of SN 2006X do not fit the computed extinguished spectra with $R_V$ = 3.1 dust (Figure~\ref{fig SN 2006X}).

\begin{table*}
\caption{KINGFISH SNe best fit results. The range is: 0.5 $>R_V>$5.5; 0.0$>A_V>$3.5, step size=0.1}
\vspace{-5mm}
\label{table KINGFISH bestfit}
\begin{center}
\begin{tabular}{lc|ccccc|cccc}
\hline\hline
   &       &      \multicolumn{5}{|c|}{\sc Best fit} & \multicolumn{4}{c}{\sc Fixed $R_V$=3.1}\\
\sc SN name & \sc $Epoch$ & \sc $R_V$ & \sc $A_V$ & \sc $E(B-V)_{Host}$ & \sc $B_{mag}$ & \sc $B-V$ & \sc $A_V$ & \sc $E(B-V)_{Host}$ & \sc $B_{mag}$ & \sc $B-V$\\ 
        &  $ (day) $ &  & (mag) & (mag)& \sc $(mag)$ & (mag) & \sc (mag) & (mag) & (mag) & (mag) \\ 
\hline\hline
\multirow{8}{*}{SN 1989B}  & 0 & 2.2 & 1.1 & 0.48 & 12.48 & 0.48 & 1.2 & 0.37 & 12.46 & 0.37\\ 
 & 6 & 2.5 & 1.0 & 0.37 & 12.47 & 0.51 & 1.0 & 0.29 & 12.39 & 0.44\\ 
 & 11 & 1.1 & 0.9 & 0.75 & 13.18 & 1.07 & 1.1 & 0.32 & 12.93 & 0.64\\ 
 & 21 & 2.9 & 1.1 & 0.32 & 14.03 & 1.17 & 1.1 & 0.30 & 14.0 & 1.15\\  
 & 22 & 5.4 & 0.9 & 0.13 & 13.73 & 1.04 & 0.9 & 0.24 & 13.85 & 1.15\\ 
 & 32 & 5.5 & 0.7 & 0.09 & 14.26 & 1.20 & 0.6 & 0.15 & 14.22 & 1.26\\ 
 & 52 & 2.5 & 0.8 & 0.27 & 15.17 & 1.19 & 0.8 & 0.21 & 15.11 & 1.13\\ \hline
\multirow{6}{*}{SN 2002bo}  & -4 & 3.1 & 1.2 & 0.37 & 14.09 & 0.29 & 1.2 & 0.37 & 14.09 & 0.29\\ 
 & -3 & 3.0 & 1.2 & 0.39 & 14.04 & 0.33 & 1.2 & 0.37 & 14.03 & 0.31\\ 
 & -1 & 2.3 & 1.2 & 0.51 & 14.11 & 0.48 & 1.3 & 0.41 & 14.10 & 0.38\\ 
 & 4 & 2.7 & 1.3 & 0.46 & 14.25 & 0.54 & 1.3 & 0.40 & 14.18 & 0.48\\   
 & 28 & 5.1 & 1.8 & 0.31 & 16.83 & 1.43 & 1.7 & 0.47 & 16.9 & 1.59\\ \hline
\multirow{8}{*}{SN 2006X}  & -6 & 1.4 & 2.0 & 1.37 & 15.66 & 1.29 & 2.2 & 0.70& 15.17 & 0.62 \\ 
 & 0 & 1.6 & 2.2 & 1.31 & 15.46 & 1.30 &  2.4 & 0.76& 15.10 & 0.75 \\ 
 & 2 & 0.9 & 1.4 & 1.45 & 14.83 & 1.48 & 1.7 & 0.53& 14.19 & 0.56 \\ 
 & 6 & 1.4 & 2.1 & 1.39 & 15.64 & 1.52 & 2.3 & 0.71& 15.15 & 0.84 \\ 
 & 8 & 1.8 & 2.7 & 1.38 & 16.37 & 1.57 & 2.8 & 0.86& 15.94 & 1.05 \\  
 & 12 & 1.3 & 2.2 & 1.49 & 16.37 & 1.84 & 2.5 & 0.75& 15.91 & 1.10 \\ 
 & 13 & 1.8 & 2.8 & 1.37 & 16.95 & 1.76 & 3.0 & 0.88& 16.65 & 1.28 \\ 
\hline\hline
\end{tabular}
\end{center}
\end{table*}

\section{Results \& discussion}

\subsection{Color excess probabilities}

As described in $\S$\ref{kfsample}, we divided the galaxies into six groups according to their morphology. For each group, we binned the color excess values and galactocentric distances from the Monte Carlo simulation in bins of 0.025 mag for E(B-V) and 0.025 for normalized galactrocenric distances $R/R_{25}$. The color excess probability maps are shown in Figure~\ref{fig all_colorplots_3x2}. The color indicates the number of Ia SNe with a certain E(B-V) value at a certain distance from the galaxy center. The variation in dust mass distribution and amount among galaxy types affects  the color excess E(B-V). From our simulation, we expect that the most extinguished SNe will occur in Sab-Sb galaxies, the most dust rich galaxy class, while SNe in S0 and Irregulars are less affected by dust.
The blue lines are the mean, i.e. most probable, E(B-V). The blue crosses are observed SNe Ia from \citet{Wang2006ApJ...645..488W} (Table 1 and 2).

\begin{figure}
\begin{center}
\includegraphics[trim=15mm 20mm 15mm 30mm, width=9cm, clip=true]{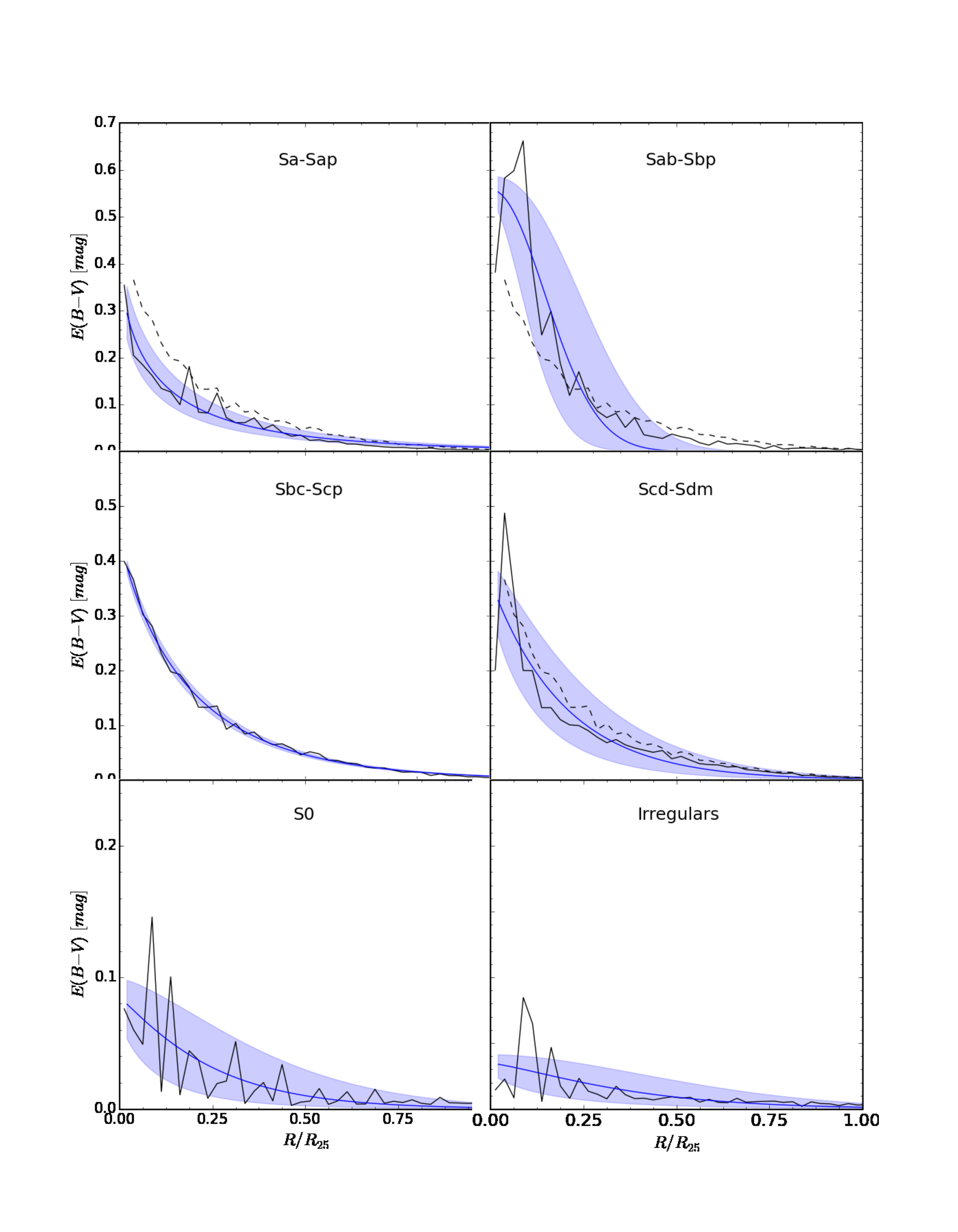}
\vspace{-5mm}
\caption{Mean E(B-V) values from our Monte Carlo simulation (black line) compared to the fitted $E_{B-V}(R/R_{25})$ functions (eq.\ref{functionebveq}, blue line) for different morphological classes. The shaded blue shows the 1-$\sigma$ uncertainty of the best fit parameters (Table~\ref{table summarized colorplot}). The dashed line in the late-type galaxies panels is the mean E(B-V) value for the Sbc-Scp galaxies group, plotted for comparison reasons.}
\label{fitted}
\end{center}
\end{figure}

We fit the mean color excess values as a function of $R/R_{25}$:
\begin{equation}
\label{functionebveq}
\mathrm{ E_{B-V}(R/R_{25}) = \textit{f}(R/R_{25}) = a\exp(b (R/R_{25})^c ) } ,
\end{equation}
where a, b and c are free parameters determined by the least square method. Figure~\ref{fitted} compares the mean E(B-V) to the fitted functions. The parameters and standard deviations for different galaxy types are given in Table~\ref{table summarized colorplot}. The most probable extinction $A_V$ can be derived from the color excess functions simply by multiplying by 3.1.\\

In $\S$3.2 we compared the observed spectra of three individual SNe Ia in three KINGFISH galaxies to our model where we assume a dust model with $R_V$ = 3.1 and uniformly distributed dust along the line of sight in the host galaxy. The color excess then depends on the depth of the supernova into the galaxy, which we take as a free random parameter (\textit{ext}). For each position there is therefore a range of possible E(B-V) values constrained by an upper limit. 

To test the reliability of our Monte Carlo simulation, we compared our calculated upper limits of E(B-V) as a a function of galactocentric distance for different morphological galaxy classes to the observed E(B-V) values of a larger sample of 109 low redshift ($\lesssim$ 0.1) SNe Ia, collected by \citet{Wang2006ApJ...645..488W}, for which they determined the reddening distribution of SNe in their respective host galaxies (Figure~\ref{fig all_colorplots_3x2}).

The color excess of all 7 SNe in Sa-Sap galaxies and 1 SN in an Irregular galaxy is lower than the upper E(B-V) limit in our Monte Carlo simulation,  consistent with the simulation. 

In Sab-Sbp galaxies, 4 of 21 observed SNe (SN 1999cl, SN 1995E, SN 2000ce and SN 1996Z) have  $\sim$0.1-0.25 mag higher E(B-V) than our simulation's upper limit. In the case of SN 1999cl, \citet{2006AJ....131.1639K} and \citet{Wang2006ApJ...645..488W} find $R_V \approx$ 1.5, indicating nonstandard dust that explains the offset.

Of the 34 SNe Ia in Sbc-Scp galaxies, only SN 1996ai is highly reddened, and whose observed E(B-V) = 1.69 $\pm$ 0.06 mag is about $\sim$0.7 mag redder than our model upper limit.

In Scd-Sdm galaxies, one of the 7 observed SNe is inconsistent with the simulation result. SN 1997br is a spectroscopically peculiar SN 1991T-like supernova, whose E(B-V) is $\sim$0.2 mag higher than the highest possible expected value in our model. \citet{Wang2006ApJ...645..488W} suggest that SN 1997br's host galaxy dust is very different from typical Milky Way dust, which may explain the inconsistency.

Finally, only 2 of 17 SNe Ia in S0 galaxies have larger reddening than expected from our Monte Carlo simulations. SN 1986G's value of  E(B-V) is about 0.25 mag  higher than the maximum expected value, and SN 1993ag, has an observed E(B-V) $\sim$ 0.1 mag, slightly larger than the upper limit. 

In total, $\sim$91$\%$ of observed SNe have a smaller color excess than the upper limit of E(B-V) in Monte Carlo, implying that our model's amount of dust is in most cases sufficient to produce the observed color excess.
The remaining 9$\%$ SNe have larger observed color excess compared to the expected model upper limit because our model underestimates either the amount of dust, or $R_V$ $<  3.1$ (e.g. SN 1999cl, SN 1997br, SN 2006X), or more likely, there is a local dust overdensity in front of the SNe (e.g. SN 2002bo).

\begin{table}
\caption{Color excess function parameters}
\vspace{-5mm}
\label{table summarized colorplot}
\begin{center}
\begin{tabular}{l p{1.6cm} p{1.6cm} l }
\hline\hline 
 &  \multirow{1}{*}{\sc \hspace{1mm} $E_{B-V}(R/R_{25}) = a\exp(b (R/R_{25})^c )$  }
 \\  \cline{2-4}
\sc Galaxy type & \sc $a$ & \sc $b$ & \sc $c$ \\ 
\hline
Sa-Sap     & 0.43(0.04)  &  -3.87(0.16)  &  0.59(0.05)  \\
Sab-Sbp    & 0.56(0.03)  &  -24.91(9.98) &  2.02(0.28) \\ 
Sbc-Scp    & 0.45(0.01)  &  -4.1(0.07)   &  0.85(0.02)  \\
Scd-Sdm    & 0.36(0.04)  &  -4.98(0.8)   &  1.00(0.16)  \\
S0         & 0.08(0.01)  &  -4.6(1.36)   &  1.11(0.32)  \\
Irregular  & 0.03(0.01)  &  -3.34(1.07)  &  1.30(0.48) \\
\hline\hline
\end{tabular}
\end{center}
\end{table}

\subsection{Estimating $R_V$ of SNe Ia host galaxy dust}

In section 4.1, we showed that starting with integrated dust mass values and $R_V$ = 3.1, we can calculate $A_V$ and determine the most likely value of E(B-V). In this section we invert the process to determine the most likely absorption-to-reddening ratio $R_V$, given E(B-V).

From the statistical sample of color excesses $E(B-V)_{SIM}$ as a function of galactocentric distance generated by the Monte Carlo simulation, we estimate the $R_V$ for host galaxy dust of the \citet{Wang2006ApJ...645..488W} sample. We grouped these SNe by host galaxy morphology as per the KINGFISH galaxies, resulting in 7, 21, 34, 7, 17 and 1 galaxies classified as Sa-Sap, Sab-Sbp, Sbc-Scp, Scd-Sdm, S0 and Irregulars respectively.

The color excess E(B-V) = $A_V$/$R_V$ is measured by assuming an intrinsic SNe brightness and a distance. $A_V$ depends on the dust column density and $R_V$ on the physical properties of dust.

If we now assume that the most probable color excess caused by dust extinction in galaxies can be described with functions given in Table~\ref{table summarized colorplot}, for a fixed extinction $A_V$, i.e. fixed dust amount, varying $R_V$ changes the color excess E(B-V). Thus, we use the least square method to determine a factor at which our functions of most probable color excess (Table~\ref{table summarized colorplot}) best fit the \citet{Wang2006ApJ...645..488W} sample. We then estimate the $R_V$ of the sample by dividing 3.1 with the determined factor.\\

\citet{Wang2006ApJ...645..488W} calculated host galaxy reddening by averaging E(B-V) determined from the (B-V) color curve at late times (cf. \citealp{Lira2005}) and E(B-V) determined from (B-V) 12 days after maximum, $\Delta C_{12}$.  SNe absolute V magnitudes were then calculated  as a function of $\Delta C_{12}$, and plotted against E(B-V) of the host galaxies. The slope is $R_V$. They derive $R_V$ = 2.30 $\pm$ 0.11, and suggest that the remarkably small scatter indicates that the dust properties are similar in z $\lesssim$  0.1 host galaxies.\\

\begin{figure}
\begin{center}
\includegraphics[trim=7mm 0mm 7mm 0mm, width=9cm, clip=true]{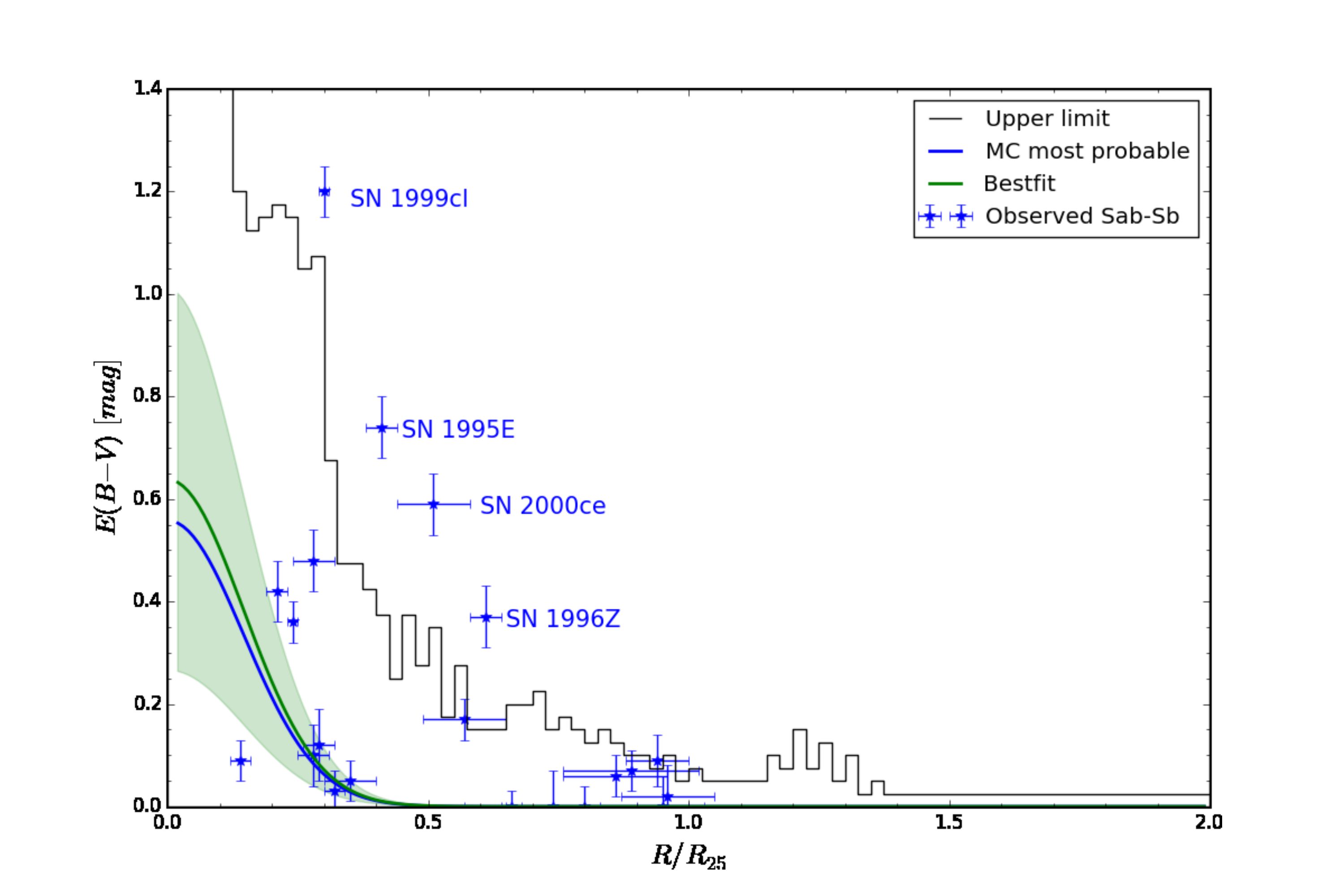}
\vspace{-5mm}
\caption{Blue crosses show a sample of 21 Ia SNe in Sab-Sbp galaxies. The sample best matches the function of most probable color excess for SNe Ia in Sab-Sbp galaxies (blue line) if multiplied by 1.14 $\pm$ 0.67 (green line). The shaded green is the 1-$\sigma$ uncertainty.}
\label{fig:fittedSb}
\end{center}
\end{figure}

\begin{figure}
\begin{center}
\includegraphics[trim=7mm 0mm 7mm 0mm, width=9cm, clip=true]{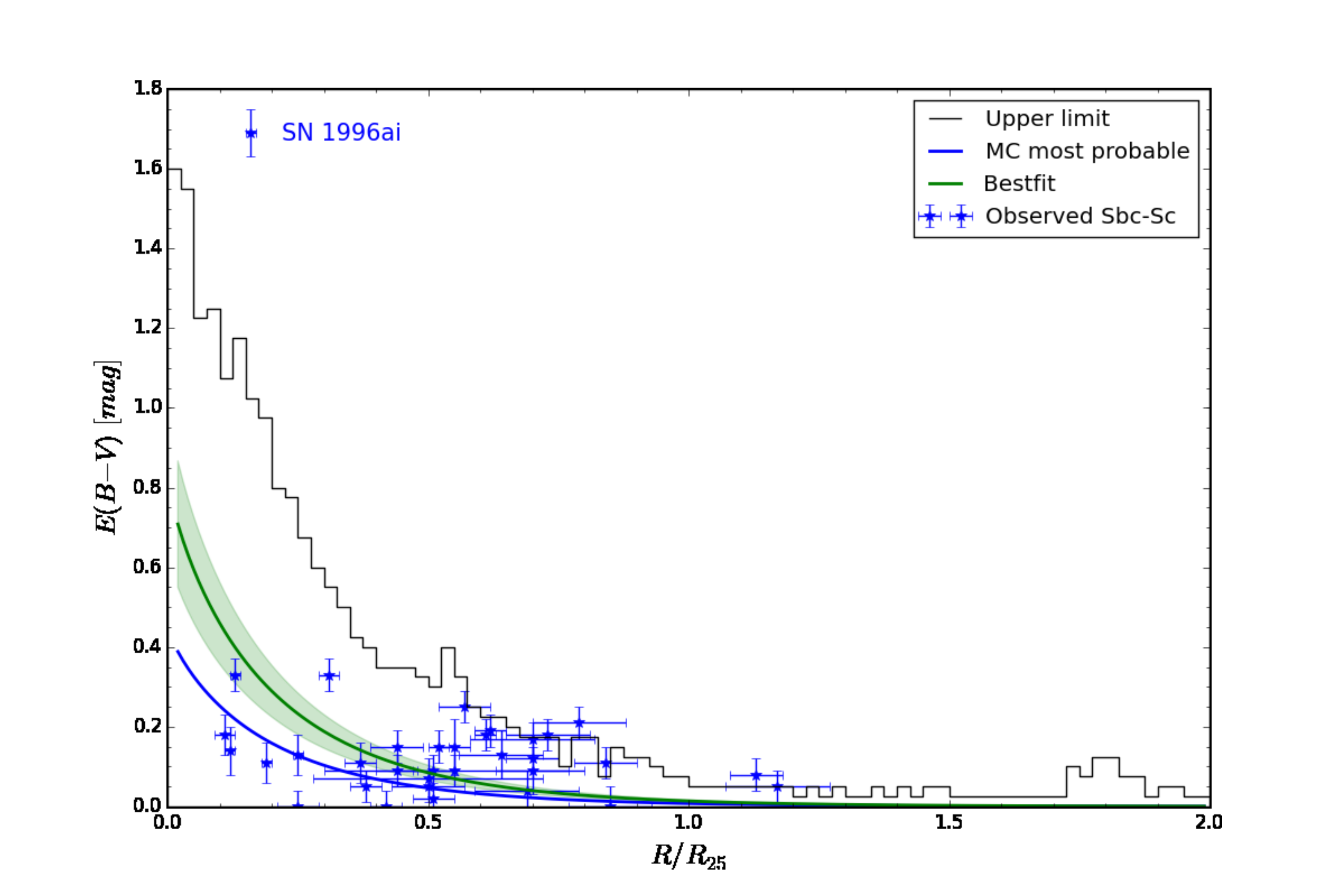}
\vspace{-5mm}
\caption{Blue crosses show a sample of 34 Ia SNe in Sbc-Scp galaxies from \citet{Wang2006ApJ...645..488W}. The sample best matches the function of most probable color excess for SNe Ia in Sbc-Scp galaxies (blue line) if multiplied by 1.82 $\pm$ 0.41 (green line). The shaded green is the 1-$\sigma$ uncertainty.}
\label{fig:fittedSc}
\end{center}
\end{figure}

We use our reddening statistics to estimate the $R_V$ value for the \citet{Wang2006ApJ...645..488W} subsamples of 21 Ia SNe observed in Sab-Sbp galaxies, and 34 SNe in Sbc-Scp. We find that our function of galactocentric distance for most probable color excess E(B-V) best matches the sample of observed SNe in Sab-Sbp galaxies if multiplied with 1.14 $\pm$ 0.67 (Figure~\ref{fig:fittedSb}), which corresponds to $R_V$ = 2.71 $\pm$ 1.58. In the case of SNe in Sbc-Scp galaxies, we find $R_V$ = 1.70 $\pm$ 0.38 (Figure~\ref{fig:fittedSc}). Both $R_V$ values are smaller than the common ratio for Milky way dust, which is consistent with the recent earlier studies (see Table~\ref{rvstudies}). This method of $R_V$ estimation requires a large sample of observed SNe. The subsample of 7, 7, 17 and 1 observed SNe in Sa-Sap, Scd-Sdm, S0 and Irregulars respectively, is not enough to get a reasonable result. \\

\subsection{Model uncertainty and effect on results}

The main uncertainty which can significantly affect the results is the systematic uncertainty of dust masses.
Our integrated dust masses are consistent with the masses calculated by \citet{2011ApJ...738...89S}, but there is a discrepancy with dust masses determined by \citet{Draine2007ApJ...663..866D} and Gordon (2008), who use different methods.
Gordon (2008) estimated the masses from 70 $\mu$m to 160 $\mu$m flux ratio, assuming the dust radiates as a black body with emissivity $\propto \lambda^{-2}$, a simple dust model consisting of 0.1 $\mu$m silicon grains. \citet{Draine2007ApJ...663..866D} calculated the masses by fitting a multicomponent dust model to the galaxy spectral energy distribution.

For each galaxy in our sample (Table~\ref{Kingfishgalaxies}) we computed dust mass ratios of the Draine, Gordon and Skibba values. The distribution is shown in Figure~\ref{fig:masshisto}.
Our dust masses, are in average 2.02 $\pm$ 1.07 and 3.12 $\pm$ 2.24 times lower than the masses derived using Karl Gordon's dust mass maps, and the masses determined by \citet{Draine2007ApJ...663..866D} respectively. The median of the ratios of Gordon's to our, and Draine's to our masses are 1.8, and 2.4 respectively. 
The discrepancy in the dust mass may be caused by different dust temperature estimates, dust models, but are also  correlated with the distance to the galaxies. We used distances given in \citet{2011ApJ...738...89S} which they took from \citet{2011PASP..123.1347K}, while \citet{Draine2007ApJ...663..866D} use distances from \citet{Kennicutt2003PASP..115..928K}. Calculating the dust masses with distances from \citet{Kennicutt2003PASP..115..928K} give us 2.7 $\pm$ 1.3, and 1.8 $\pm$ 0.7 times lower masses from \citet{Draine2007ApJ...663..866D} and Gordon respectively. The dust mass ratios are summarized in Table~\ref{table DM ratios}.
 
The total dust mass is linearly related to $A_V$ and E(B-V), and thus the color excess functions given in Table~\ref{table summarized colorplot}. If there would hypothetically be twice as much dust in our model, the color excess would be twice as large, and the estimated $R_V$ ratios would be larger; $R_V$ = 3.40 $\pm$ 0.76 for Sab-Sb, and $R_V$ = 5.42 $\pm$ 3.16 for Sbc-Sc.

\begin{table}
\caption{Dust mass ratio comparison}
\vspace{-5mm}
\label{table DM ratios}
\begin{center}
\begin{tabular}{l || l l | l l}
\hline\hline 
Distances & \multicolumn{2}{c|}{ \citet{2011ApJ...738...89S} } & \multicolumn{2}{c}{ \citet{Kennicutt2003PASP..115..928K}} 
 \\  \cline{1-5} 
  Ratio    & Mean & Median  &  Mean & Median  \\
\hline
Our/Skibba & 1.43(1.35) & 1.18  & 1.54(1.38) & 1.20 \\
Draine/Skibba & 3.81(2.70) & 2.73  & \ldots & \ldots \\
Gordon/Skibba & 2.73(2.98) & 1.97  & \ldots & \ldots \\
Draine/Our & 3.12(2.24) & 2.35 & 2.73(1.31) & 2.46 \\
Gordon/Our & 2.02(1.07) & 1.82 & 1.82(0.72) & 1.74 \\
\hline\hline
\end{tabular}
\end{center}
\end{table}

\subsection{Effects of circumstellar dust} 

In this work we assume that extinction is produced only by dust in the SN host galaxy's ISM,  although there is evidence that circumstellar material (CSM), ejected by the progenitor system, may  contribute to normal SNe Ia reddening.

\citet{Patat2007Sci...317..924P} performed high resolution spectroscopy of the highly reddened SN 2006X from five different epochs and observed clear time evolution of the Na I D doublet lines. They suggest that the evolution is caused by changes in CSM ionization conditions induced by the variable SN radiation field. They calculated the Na I column density from the most intense feature at day +14, N(Na I) $\cong$ $10^{12}$ $cm^{-2}$. For comparison, we calculated the extinction $A_V \approx$ 0.001 mag, assuming the solar Na abundance,  log Na/H = -6.3, and the attenuation per unit column density for standard interstellar dust as used in our model, $A_V/N_H$ = 5.3$\times$ 10$^{-22}$ mag cm$^{2}$ H$^{-1}$ \citep{Draine2007ApJ...657..810D}.

\citet{2009ApJ...693..207B} analyzed low resolution, high signal to noise spectra of 294 SNe Ia obtained by the CfA Supernova Program, and were able to measure changes in the equivalent width of the NaI lines in two SNe: SN 1999cl and SN 2006X. The deduced Na I column density is log N(Na I) $\approx$ 14.3, corresponding to $A_V \approx$ 0.21 mag, assuming the \citep{Draine2007ApJ...657..810D} dust model and solar Na abundance. Like SN 2006X, SN 1999cl is also highly reddened:  E(B-V) $\approx$ 1.2, $R_V \approx$ 1.6. This detection of variable Na I D features in two of the most reddened SNe suggests that the change in EW with time might be associated with circumstellar or interstellar absorption \citep{2009ApJ...693..207B}. However, no such variation is detected in the highly reddened SN 2003cg (E(B-V) $\approx$ 1.3 mag, $R_V \approx$ 1.8) \citep{2006MNRAS.369.1880E}. 

\citet{2007Sci...315..212W} found a strong correlation between polarization across the Si II line and $\bigtriangleup m_{15}$, possibly indicative of large-scale asymmetries, such as large plumes located  above the SN photosphere, or generated from the interaction of the ejecta with circumstellar material, such as an accretion disk before the explosion of the white dwarf progenitor.

\citet{2013MNRAS.431L..43J} used Herschel PACS 70 $\mu$m observations of three nearby SNe Ia to search for pre-existing circumstellar dust. They derive upper limits on CS dust masses based on nondetection. Assuming 0.1 $\mu$m graphitic or silicate dust grains in a pre-existing dust shell with radius $r_d$ $\sim$ 10$^{17}$ cm (cf. \citet{Patat2007Sci...317..924P}), heated to $T_d$ $\sim$ 500 K, they obtained $M_d \leq$ 7 $\times$ 10$^{-3}$ M$_{\odot}$ for SN 2011fe, and $M_d$ $<$ 10$^{-1}$ $M_{\odot}$ for SNe 2011by and 2012cg. Assuming that the dust is distributed in a dust shell of $r_d \sim$ 10$^{17}$ cm, we can exclude dust mass surface densities $\sigma_{dust}$ $>$ 10$^{-4}$  g/cm$^2$ and $\sigma_{dust} \gtrsim$ 1.6 $\times$ 10$^{-3}$ g/cm$^2$ for SN 2011fe and for SNe 2011by and 2012cg respectively. Those upper limits are still at least an order of magnitude larger than the dust mass surface densities required to produce significant extinction, so that it can not be ruled out that there is CSM, which affects the colors Ia SNe.

At present, the available studies can neither definitely exclude SNe Ia reddening by circumestellar dust nor affirm that CSM dust is responsible for the low $R_V$ values.\\

\section{Summary}

\begin{itemize}
\item  We use dust temperature estimations by \citet{2011ApJ...738...89S} and Herschel 500 $\mu$m SPIRE maps to create pixel-by-pixel dust mass surface density maps for a sample of 59 KINGFISH galaxies \citep{2011PASP..123.1347K}.
\item Comparison of our model to SN 1989B, SN 2002bo and SN 2006X, three individual SNe Ia which occurred in KINGFISH galaxies, shows that our model is in good agreement with the observations. We are able explain and reproduce the observed spectra. 
\item We grouped the KINGFISH sample into six galaxy groups according to the morphology, and ran Monte Carlo simulations to place constrains on the color excess as a function of galactocentric distance. 
We developed a color excess probability model due to dust extinction and show the differences between galaxy groups. We find that the largest reddening probability is expected in Sab-Sbp and Sbc-Sc galaxies, while S0 and Irregulars are very dust poor. We determine the most probable reddening for different galaxy classes as a function of $R/R_{25}$ (Table~\ref{table summarized colorplot}). The functions can be used to estimate reddening of SNe Ia depending on the host galaxy morphology and the galactocentric distance of the observed SNe, or for extinction correction studies for Type Ia supernova rates (e.g. \citet{2005MNRAS.362..671R}).
\item We present a new approach for determination of the absorption-to-reddening ratio $R_V$ using statistics of color excess developed in a Monte Carlo simulation. We find for a sample of 21 Ia SNe observed in Sab-Sbp galaxies, and 34 SNe in Sbc-Scp, an $R_V$ of 2.71 $\pm$ 1.58 and $R_V$ = 1.70 $\pm$ 0.38 respectively, but the results strongly depend on the derived values for dust mass.\\
\end{itemize}

\begin{acknowledgements}
Acknowledgements. This work was supported by the Space Telescope Science Institute Summer Student Program (SASP) and the STScI Director's Discretionary Research Fund. We thank Eric Hsiao  for reading the draft paper and giving comments, Karl Gordon for providing his SINGS galaxies dust mass maps, and Norbert Przybilla, Bruce Draine, Andrew Howell, Nick Mostek and Andy Fruchter for helpful discussions and advice.

Figures~\ref{figHSTSN1989B} and ~\ref{ngc3190}, use images based on observations made with the NASA/ESA Hubble Space Telescope, and obtained from the Hubble Legacy Archive, which is a collaboration between the Space Telescope Science Institute (STScI/NASA), the Space Telescope European Coordinating Facility (ST-ECF/ESA) and the Canadian Astronomy Data Centre (CADC/NRC/CSA).

Figure~\ref{figHSTSN1989B} uses a SDSS image of NGC 3627.
Funding for the SDSS and SDSS-II has been provided by the Alfred P. Sloan Foundation, the Participating Institutions, the National Science Foundation, the U.S. Department of Energy, the National Aeronautics and Space Administration, the Japanese Monbukagakusho, the Max Planck Society, and the Higher Education Funding Council for England. The SDSS Web Site is http://www.sdss.org/.

The SDSS is managed by the Astrophysical Research Consortium for the Participating Institutions. The Participating Institutions are the American Museum of Natural History, Astrophysical Institute Potsdam, University of Basel, University of Cambridge, Case Western Reserve University, University of Chicago, Drexel University, Fermilab, the Institute for Advanced Study, the Japan Participation Group, Johns Hopkins University, the Joint Institute for Nuclear Astrophysics, the Kavli Institute for Particle Astrophysics and Cosmology, the Korean Scientist Group, the Chinese Academy of Sciences (LAMOST), Los Alamos National Laboratory, the Max-Planck-Institute for Astronomy (MPIA), the Max-Planck-Institute for Astrophysics (MPA), New Mexico State University, Ohio State University, University of Pittsburgh, University of Portsmouth, Princeton University, the United States Naval Observatory, and the University of Washington.

This research made use of APLpy, an open-source plotting package for Python hosted at http://aplpy.github.com.

\end{acknowledgements}

\bibliographystyle{apj}
\bibliography{ref2}

\end{document}